\documentclass[journal,letterpaper]{IEEEtran}
\usepackage{amsmath}
\usepackage{amsthm}
\usepackage{amsfonts}
\usepackage{amssymb}
\usepackage{makeidx}
\usepackage{cite}
\usepackage{graphicx,bigints}
\usepackage{mathtools}
\usepackage{cases}
\usepackage{color}
\usepackage{hyperref}
\usepackage{bbm}
\usepackage[normalem]{ulem}
\usepackage{braket}
\usepackage{textcomp,gensymb}

\usepackage{enumitem}  

\usepackage{epstopdf}
\usepackage{xcolor}
\usepackage{lipsum}

\usepackage{multirow}

\usepackage[makeroom]{cancel}

\makeatletter
\DeclareFontFamily{U}{tipa}{}
\DeclareFontShape{U}{tipa}{m}{n}{<->tipa10}{}
\newcommand{\arc@char}{{\usefont{U}{tipa}{m}{n}\symbol{62}}}%

\newcommand{\arc}[1]{\mathpalette\arc@arc{#1}}

\newcommand{\arc@arc}[2]{%
	\sbox0{$\m@th#1#2$}%
	\vbox{
		\hbox{\resizebox{\wd0}{\height}{\arc@char}}
		\nointerlineskip
		\box0
	}%
}
\makeatother

\newtheorem{theorem}{Theorem}
\newtheorem{corollary}{Corollary}
\newtheorem{lemma}{Lemma}

\newtheorem{proposition}{Proposition}
\newtheorem{remark}{Remark}

\DeclareMathOperator{\cE}{\mathcal{E}}
\DeclareMathOperator{\cO}{\mathcal{O}}
\DeclareMathOperator{\cI}{\mathcal{I}}
\DeclareMathOperator{\cL}{\mathcal{L}}

\DeclareMathOperator{\bS}{\mathbb{S}}

\DeclareMathOperator{\cR}{\mathcal{R}}

\DeclareMathOperator{\SINR}{\textnormal{SINR}}

\DeclareMathOperator{\bR}{\mathbb{R}}

\DeclareMathOperator{\bP}{\mathbf{P}}
\DeclareMathOperator{\ind}{\mathbbm{1}}
\DeclareMathOperator{\bE}{\mathbf{E}}
\DeclareMathOperator{\bZ}{\mathbb{Z}}
\newcommand*\diff{\mathop{}\!\mathrm{d}}

\newcommand*\nnb{\nonumber}

\newcommand\independent{\protect\mathpalette{\protect\independenT}{\perp}}
\def\independenT#1#2{\mathrel{\rlap{$#1#2$}\mkern2mu{#1#2}}}

\newcommand{\ea}{\stackrel{(\text{a})}{=}}
\newcommand{\eb}{\stackrel{(\text{b})}{=}}
\newcommand{\ec}{\stackrel{(\text{c})}{=}}
\newcommand{\ed}{\stackrel{(\text{d})}{=}}

\definecolor{sandy}{HTML}{E6E2AF}
\definecolor{stone}{HTML}{A7A37E}
\definecolor{beach}{HTML}{EFECCA}
\definecolor{ocean}{HTML}{046380}
\definecolor{diver}{HTML}{002F2F}

\definecolor{Firenze1}{HTML}{468966}
\definecolor{Firenze2}{HTML}{FFF0A5}
\definecolor{Firenze3}{HTML}{FFB03B}
\definecolor{Firenze4}{HTML}{B64926}
\definecolor{Firenze5}{HTML}{8E2800}
\definecolor{mediumpersianblue}{rgb}{0.0, 0.4, 0.65}
\definecolor{hongik}{HTML}{004498}
\definecolor{cobalt}{rgb}{0.0, 0.14, 0.86}
\definecolor{burntorange}{rgb}{0.8, 0.33, 0.0}

\definecolor{ultramarineblue}{rgb}{0.25, 0.4, 0.96}
\newcommand{\cc}[1]{\textcolor{black}{#1}}

\title{A Novel Analytical Model for LEO and MEO Satellite Networks based on Cox Point Processes}

\author{Chang-Sik Choi and François Baccelli
	\IEEEcompsocitemizethanks{\IEEEcompsocthanksitem{Chang-Sik Choi is with Hongik University, South Korea. (email: chang-sik.choi@hongik.ac.kr). François Baccelli is with Inria Paris and Telecom Paris, France.  (email: francois.baccelli@inria.fr). 	The work of Chang-Sik Choi was supported by
		the NRF RS-2024-00334240. 
		The work of Francois Baccelli was supported by the ERC NEMO
		grant 788851 to INRIA  and by the French National Agency for Research via the project n°ANR-22-PEFT-0010 of the France 2030 program PEPR réseaux du Futur.}}
}
\begin{document}
	\maketitle 
\begin{abstract}
\cc{This work develops an analytical framework for downlink low Earth orbit (LEO) or medium Earth orbit (MEO) satellite communications, leveraging tools from stochastic geometry. We propose a tractable approach to the analysis of such satellite communication systems,  accounting for the fact that satellites are located on circular orbits. We accurately incorporate this geometric property of LEO or MEO satellite constellations by developing a Cox point process model that jointly produces orbits and satellites on these orbits. Our work contrasts with previous modeling studies that presumed satellite locations to be entirely random, thereby overlooking the fundamental fact that satellites are jointly positioned on orbits. Employing this Cox model, we analyze the network performance experienced by users located on Earth. Specifically, we evaluate the no-satellite probability of the proposed network and the Laplace transform of the interference created by such a network. Using it, we compute its SIR (signal-to-interference) distribution, namely its coverage probability. By presenting fundamental network performance as functions of key parameters, this model allows one to assess the statistical properties of downlink LEO or MEO satellite communications and can thus be used as a system-level design tool to operate and optimize forthcoming complex LEO or MEO satellite networks.}
\end{abstract}
	
	
\section{Introduction}
	\subsection{Motivation and Related Work}
Satellite communications provide global-scale connectivity to users everywhere on earth without the need for deploying base stations and infrastructure on the ground \cite{8002583,8700141}. As orbiting the Earth at very high speeds, LEO and MEO satellites provide reliable and fast Internet connectivity to millions of devices \cite{8002583,8700141}. In the early stage of satellite communications, the number of deployed satellites was very small, and thus only a limited number of satellites were available for connections. For instance, the Iridium constellation  \cite{Iridium}---6 orbits with 11 satellites on each orbit---provided a call coverage of up to 7 minutes. More precisely, because of the motions of satellites and their limited numbers, the call was either dropped or transferred after about 7 minutes. Preventing outages was one of the key design criteria for satellite communications of this early stage. 
	
	In modern LEO or MEO satellite communication systems with numerous satellites, the goal is not only to prevent call outages but also to provide high-speed and low-latency Internet connections to millions of devices. \cc{Comprised of a large number of satellites\cite{9351765,10436074}, often integrated with terrestrial network infrastructures \cite{8626457,9502642}, satellite communication systems are envisioned to handle various demands of ground or even aerial devices around the globe.}  Recently, several companies intended to establish their satellite communication networks by building such large constellations for global connectivity \cite{FCCKuiper,FCCBoeing,okla}. It is not hard to imagine a large number of satellites from various companies will provide Internet connections to devices anywhere on earth. Since the spatial distribution of satellites determines the performance of communications on satellite networks, building an analytical framework is vital to the description and analysis of the satellite communications.

	To provide an analytical framework for describing the locations of LEO or MEO satellites, several recent papers \cite{9079921,9177073,9218989,9497773,9678973,9861782} used binomial or Poisson point process models in stochastic geometry \cite{chiu2013stochastic,baccelli2010stochastic,baccelli2010stochasticvol2}. The main advantage of employing these models lies in the fact that it not only provides geometric treatment to the interference and coverage in such networks but also identifies key network performance behaviors as functions of networks' distributional parameters  \cite{6042301,haenggi2012stochastic,8419219,Choi2018Densification}. \cc{To take advantage of assessing typical network performance and of getting high-level insights,  \cite{9079921,9177073,9218989,9497773,9678973,9861782} modeled the distribution of LEO or MEO satellites as binomial or Poisson point processes, where the satellites' locations are assumed to be uniformly distributed on a sphere. }  In \cite{9079921,9177073,9218989,9497773,9678973}, the impact of interference created by a binomial satellite point process was evaluated and accounted for the SINR coverage probability in \cite{9861782}. It is important to note that the recent modeling technique based on binomial or Poisson point processes \cite{9079921,9177073,9218989,9497773,9678973,9861782} overlooked the essential geometric facts that LEO or MEO satellite constellations are comprised of orbits and that they are always on those orbits. \cc{Although binomial or Poisson models provide new results on the performance metrics of downlink communications, the existence of an orbital structure and the impact of the structure to the network performance cannot be examined.} This motivates us to develop an isotropic Cox point process model that incorporates these geometric characteristics. This process jointly produces orbits and satellites on these orbits, isotropically distributed in space. This Cox model allows one to analyze the network performance seen by users anywhere on Earth. To the best of our knowledge, this is the first work that explicitly models the orbits and satellites on them and analyzes the typical network performance  of downlink communications from satellites to network users anywhere on Earth.  To demonstrate the use of the developed analytical framework, we derive the no-satellite probability and then the coverage probability of downlink communication from satellite to a typical user. \cc{It is important to note that analytical models such as the binomial and Poisson models approximate an existing or forthcoming LEO or MEO constellation by delineating the local geometry of the target constellation. Such a local approximation of the constellation is known to produce coverage probability approximating well that of the target constellation \cite{9079921,9177073,9218989,9497773}. Similar to those studies, we also show that the Cox model effectively represents an existing or forthcoming constellation by better approximating the target network's local geometry as seen by the typical user. We present numerical experiments demonstrating the use of the proposed framework.}
	

\subsection{Theoretical Contributions}
\subsubsection{Orbit-based Stochastic Geometry Model For LEO or MEO satellite constellations} 
\cc{This work develops an analytical framework for LEO or MEO satellite networks using stochastic geometry. The proposed Cox model for LEO or MEO satellites stands in contrast to most existing frameworks that have employed binomial point processes to describe the locations of satellites as random points. While recent works show that the binomial or Poisson point process models locally portray some existing satellite constellations \cite{9079921,9177073,9218989,9497773,9678973,9861782}, this approach overlooked the fact that satellites are always located on orbits, which affects network geometry and the networks therein. This Cox framework serves as one among several modeling techniques for LEO or MEO satellite networks. To demonstrate its applicability, Section \ref{S:6} shows that how the Cox model approximates an upcoming constellation in terms of the coverage probability. It also demonstrates that how the Cox point process model can reproduce LEO or MEO satellite constellations by adjusting its two parameters: the mean number of orbits and the mean number of satellites per orbit.}  
\subsubsection{Performance Analysis of Downlink Communications:} 
We first prove that our developed Cox point process is isotropic, namely invariant by all rotations. Leveraging this rotation invariance property, we derive a closed-form expression for the arc length of an orbit and provide an integral formula for the distribution of the distance from a typical user to its nearest satellite. These expressions are essential in the derivation of the network performance of LEO or MEO satellite downlink communications. For example, leveraging this, we derive the probability that a typical user has no visible satellite as a function of $ \lambda $ the orbit density, $ \mu $ the density of satellites per orbit, and $ r_a $ the satellite altitude. \cc{Further, we focus on the interplay of network geometry and derive the SIR coverage probability of the typical user under Nakagami-$ m $ fading by assuming that the interference power dominates the noise. We compare the derived coverage formula to the results obtained by Monte Carlo simulations and validate the derived formulas.}

\subsubsection{Design Insights for Practical LEO and MEO satellite Networks} Our approach provides a comprehensive tool for designing and enhancing LEO or MEO satellite communication systems. For example, we derive an expression for the interference experienced by the typical user, which can be utilized to design interference management techniques in densely-deployed LEO or MEO satellite communication systems. Additionally, this paper obtains expressions for the no-satellite probability and the coverage probability as the functions of key parameters such as $\lambda$ and $\mu$, allowing us to assess each variable's individual impact on the system's large-scale performance. Leveraging the derived formula, network operators of satellite systems can explore and assess various deployment options by controlling the number of satellites per orbit or the number of orbits per altitude, without time-consuming system-level simulation. Lastly,  the proposed model and analysis can serve as a basis for assessing the time-domain performance metrics of satellite networks, such as the time fraction of coverage or delay.
%
%

	\section{System Model}
		
	\subsection{Models for Orbits and Satellites}

We denote by $ r_e $ the radius of Earth ($r_e \approxeq6400 $ km) and the center of Earth is located at the origin of the Euclidean space. The {reference plane} is the $ xy $ plane and the longitudinal zero point is the $ x $-axis. We denote by $ r_a $ the altitude of the satellites or equivalently the radius of orbits. All orbits are assumed to be circles centered at the origin.  Let $ r_s  = r_e+r_a,$ where $r_s$ is the radius of orbits.  

	\begin{figure}
	\centering
	\includegraphics[width=.8\linewidth]{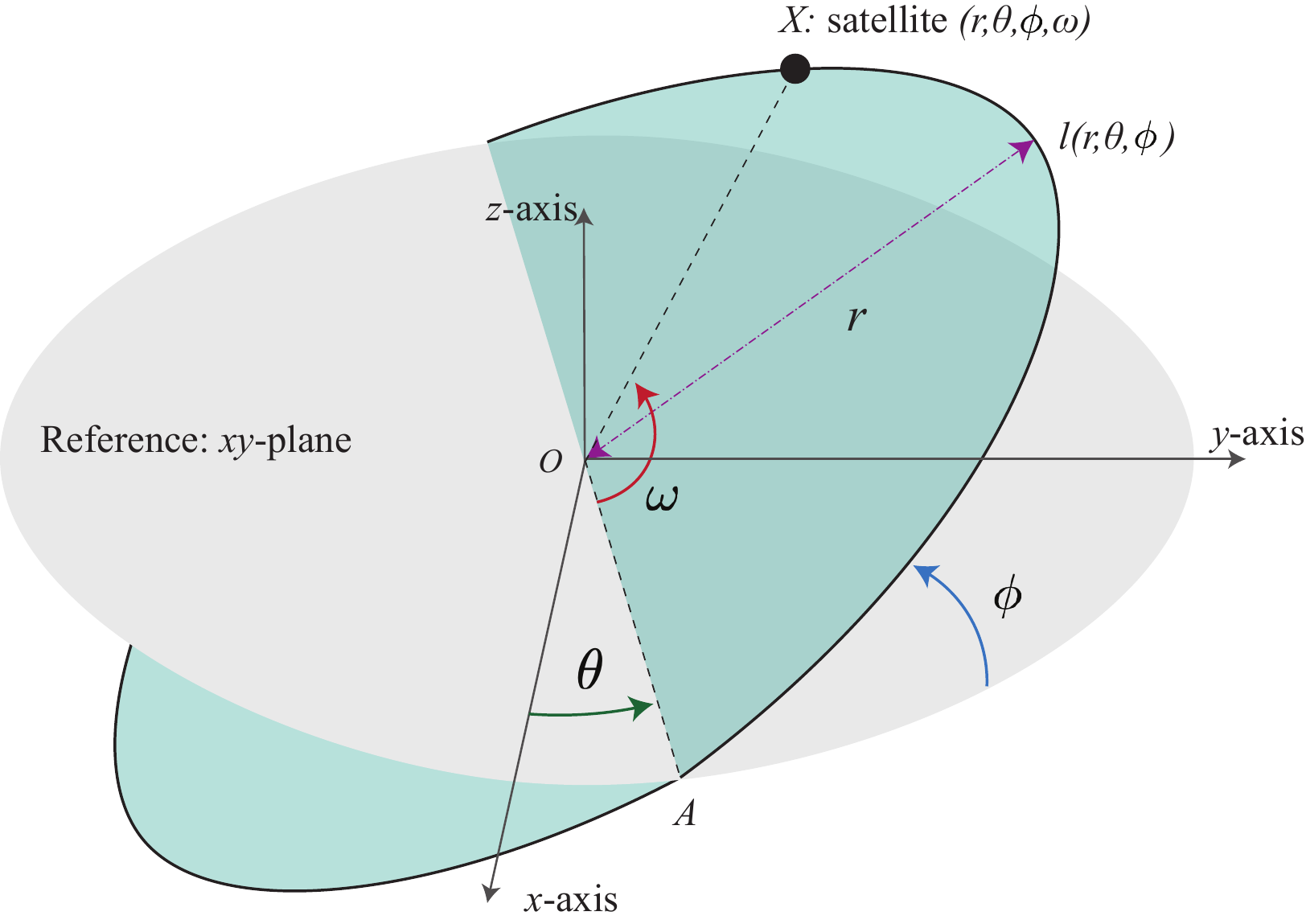}
	\caption{Let $A$ be the ascending/descending point. The longitude $ \theta $ is the angle that $ \overline{OA} $ makes with the $ x $-axis. The inclination $ \phi $ is the angle that the orbital plane makes with the reference plane.  $ \omega $ is the angle that the $ \overline{OX} $ makes with $ \overline{OA}.$ }
	\label{fig:angledefpdf}
\end{figure}

To model circular orbits of LEO or MEO satellites in the simplest case, we first consider a Poisson point process $ \Xi  $ of density $ \lambda\sin(\phi)/(2\pi) $ on the rectangle set $ \cR  {=} [0,\pi ) \times [0, {\pi} )$. We write $\Xi = \sum_{i}\delta_{(\theta_i,\phi_i)}$. Each point of the Poisson point process on $ \cR $, say $ (\theta,\phi)$, is mapped to an undirected orbit $ l(\theta,\phi) $ in the Euclidean space  $ \bR^3 $. Specifically, $ \theta $ gives the longitude of the orbit $l(\theta,\phi)$ and $ \phi $ is the inclination of the orbit $l(\theta,\phi)$. See Fig. \ref{fig:angledefpdf} for the longitude and inclination. The orbit process $ \cO $ on $\bR^3$ is 
\begin{equation}
	\cO = \bigcup_{i\in\bZ }l(\theta_i,\phi_i)=\!\!\bigcup_{{(\theta_i,\phi_i)}\in\Xi}\!\!l(\theta_i,\phi_i).
\end{equation} 
The orbit process is a union of circles located on the sphere $ \bS_{r_s}{=}\{(x,y,z)|x^2+y^2+z^2= r_s^2\} $.  Since the density of the Poisson point process $ \Xi $ is $ \lambda\sin(\phi)/(2\pi) $, there are $ \lambda $ points on $ \cR $ on average, or equivalently, there are $ \lambda $ orbits on $ \bR^3$ on average. The orbit process $\cO$ is isotropic, namely rotation invariant. We will shortly prove its rotation invariance property in Section \ref{S:properties}. 
	\par To represent the locations of satellites on orbits, we leverage the conditional structure based on a Cox point process. Specifically, conditionally on each orbit $ l(\theta_i,\phi_{i}) $, the satellites on each orbit are modeled as a Poisson point process $\psi_i$ of mean $\mu$. Because of the conditional structure, our developed satellite model geometrically ensures that the satellites are exclusively on orbits given by $\cO$.  Based on the above construction, the orbital angles $ \omega $ of satellites (See. Fig. \ref{fig:angledefpdf}.) are also characterized as a Poisson point process of intensity $ \mu/ (2\pi) $ on the finite interval $ [0,2\pi). $  
	
	Collectively, the satellite point process $ \Psi $ on $\bR^3$ is $		\Psi = \sum_{i \in \bZ} \psi_i.$ The satellite point process is defined conditionally on the orbit process and it is hence a Cox point process \cite{chiu2013stochastic,baccelli2010stochastic}. 

It is worth noting that the conditional structure of Cox point process was found to be useful in the modeling of vehicular networks on two-dimensional $\bR^2$ plane since it jointly creates road systems and vehicles on them \cite{8340239,8357962,8419219}. In the same way as vehicles are on roads in two-dimensional vehicular networks, satellites are on orbits in three-dimensional satellite networks.

\begin{figure}
	\centering
	\includegraphics[width=1\linewidth]{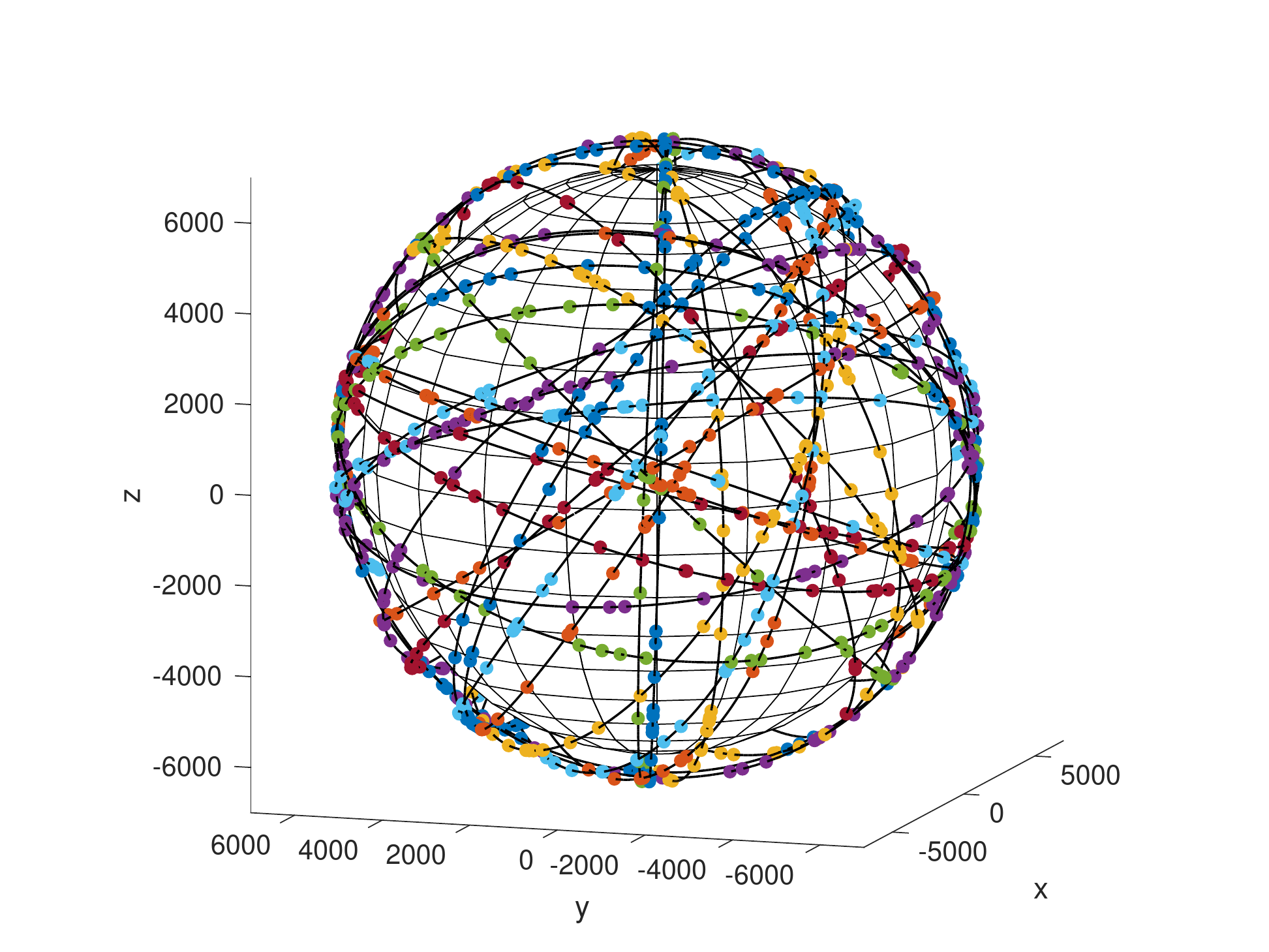}
	\caption{The proposed model with $ \lambda = 30 $, $\mu = 40 $,   and $ r_s = 7000 \text{ km}$.   }
	\label{fig:1}
\end{figure}

\begin{figure}
	\centering
	\includegraphics[width=1\linewidth]{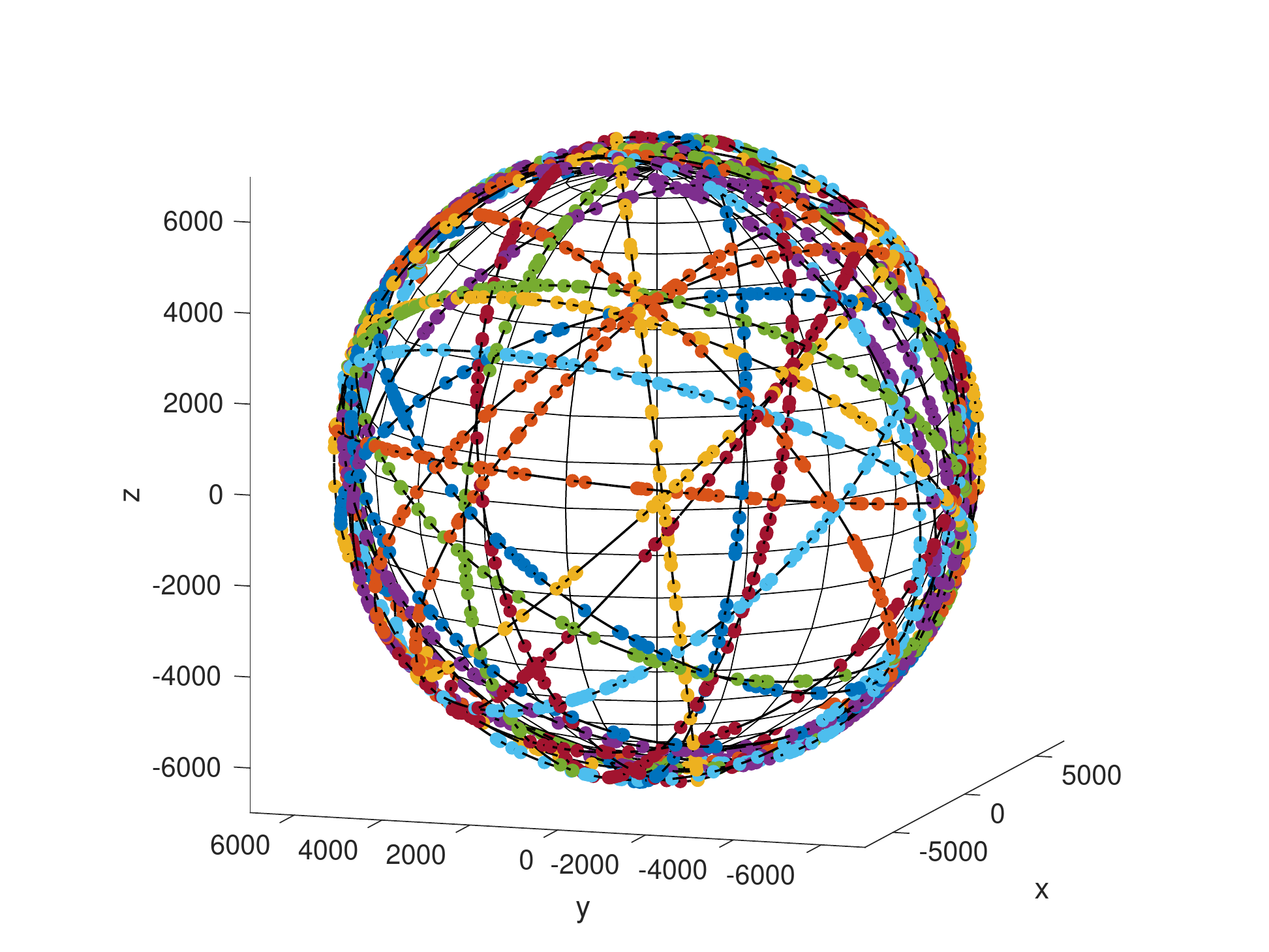}
	\caption{The proposed model with $ \lambda = 28 $, $\mu = 120$,   and $ r_s = 7000 \text{ km}$.}
	\label{fig:2}
\end{figure}

Figs. \ref{fig:1}--\ref{fig:2} illustrate the proposed network model with various $ \lambda $ and $ \mu $ parameters. The proposed stochastic geometry framework of populating orbits and their corresponding satellites is designed to easily change the number of satellites, the number of orbits, or their topological characteristics.   For instance, we may increase the total number of satellites by individually increasing $ \lambda $ or  $ \mu $ or both. Fig. \ref{fig:2} shows $\mu=120$ and since satellites are densely populated on the finite orbital planes, the clustering of the satellites on their orbits are more pronounced.

 \subsection{User Location}
\cc{In \cite{9079921,9177073,9218989,9497773,9678973,9861782}, users are assumed to be uniformly distributed on Earth. In the aforementioned work, the network users are specifically modeled as a Poisson point process of intensity $ \lambda_u/(4\pi^2r_e^2) $ on the sphere $ \{(x,y,z)\in\bR^3| x^2+y^2+z^2 = r_e^2\} $ where  $ \lambda_u $ is the number of network users. }
 
\cc{To reduce the computational complexity accompanied with the distribution of users, we also assume that users are uniformly distributed on Earth, independent of satellites. For downlink communications, users are assumed to get their downlink signals from their closest satellites \cite{38821,9079921,9177073,9218989,9497773,9678973,9861782}.} 

\cc{This nearest association assumption facilitates the analysis of SINR coverage by taking the most significant signal as the desired signal and treating the signals from other satellites as interference. in practice the nearest association can be implemented to each UE, first by detecting various signals from visible satellites and then by determining the closest satellites by time-averaging of those signals.}

\subsection{Propagation Model}
	The downlink signals from satellites attenuate because of Doppler shift, weather, rain,  reflection from objects, and path loss over the space. In practice, Doppler shift can be compensated by exploiting the existing data on satellites' orbits and speeds \cite{7468478,6184256}. A few large-scale uniform factors such as rain attenuation can be ignored under certain conditions \cite{6184256,9218989,9497773,9678973}. To emphasize the role of satellite geometry onto the network performance, we propose a simple propagation model where various attenuating and propagation factors are assumed to be aggregated to a single independent fading random variable. This approach was found useful in recent work \cite{9218989,9497773,9678973}, where analytical tractability was achieved by focusing on the impact of topology of satellite communication systems. 
		 		
	Suppose a transmitter and a user separated by a distance $ d. $ Assume the transmitter is visible to the user. We assume that the received signal power at the user is then given by 
	\begin{equation}\label{rxpower}
		p G_{1}(\zeta)G_{2}(\zeta) H d^{-\alpha},
	\end{equation}
	where $ p $ is the receive signal power at $1$ meter, $ G_{1}(\zeta) $ is the transmit satellite antenna gain, $ G_{2}(\zeta) $ is the receive antenna gain of network users, $ H $ is a random variable representing small-scale fading, and $ \alpha $ is the path loss exponent.

To make the analysis tractable, we assume that network users and satellites are able to direct their antennas toward their associated counterparts by using technologies such as phased antenna array \cite{7271009,9678973}. We assume that the gain $ G_{1}(\zeta) $ of the LEO or MEO satellites is 
	 		\begin{align}
	 			G_{1}(\zeta)=\begin{cases}
	 				g& |\zeta| < \zeta_t, \\
	 				1&|\zeta| > \zeta_t, 
	 			\end{cases}
	 		\end{align} 
 		where $ \zeta $ is the boresight angle from the antenna's maximum radiated power direction. 
 		
 		Similarly,  network users are assumed to have isotropic antennas $ G_2(\zeta)=g_r=1 $ for all $\zeta$, namely $0$ dBi \cite{38811,38821}. 
	 		\par Assuming general small-fading, we use a Nakagami-$ m $ fading \cite{9861782}.  The CCDF of the random variable $ {H} $ is 
	 		\begin{align}
	 			\bP(H>x) &= 
	 				e^{-mx}\sum\limits_{k=0}^{m-1} \frac{(m x)^k}{k!} &\forall x\geq 0,\nnb
	 		\end{align}
where $ m\geq 1 $. When $ m=1$, Nakagami-$ m $ fading becomes Rayleigh fading and $ H $ becomes an exponential random variable with mean one. 
 		
	\subsection{Performance Metrics}
			 A downlink satellite communication system is built to cover all network users. Nevertheless, some users on the surface of the earth may not be well covered because of the lack of visible satellite or weak signals. 
			 
			 We define the no-satellite event as the event that there is no visible satellite from a network user. Since the Earth rotates and the relative locations of satellites vary over time, we characterize this through the no-satellite probability.  

			\cc{Then, to evaluate the basic performance of downlink satellite communication systems, we study the coverage probability, namely the CCDF of the SINR of network users. The coverage probability incorporates the path loss, the topological properties of the association satellite and of the interfering satellites, the small-scale fading, the antenna gains, and the background thermal noise. The coverage probability of a user at an arbitrary location $ y $, or equivalently the CCDF of SINR of the user at $y$, is given by 
			\begin{align}
&	\bP(\SINR_y>\tau )\nnb\\
&=\bP_{y}\left(\frac{p G_{X_y;y}H_i \|X_y-y\|^{\alpha}}{kTB_w+\sum_{X_j\in \bar{\Psi}\setminus X_y} pG_{X_j;y}H\|X_j-y\|^\alpha}> \tau \right)\nnb\\
				&=\bP_{y}\left(\frac{p g H_i \|X_y-y\|^{\alpha}}{kTB_w+\sum_{X_j\in \bar{\Psi}\setminus X_y} p  H_j\|X_j-y\|^\alpha}> \tau \right)\label{eq:SINRdef},
			\end{align}}
\cc{where $ X_y $ is the location of the satellite that serves the user at $ y $, $ G_{X_i;y} $ is the aggregate antenna gain from the satellite at $ X_i $ toward the user at $ y $, $ kTB_w $ is the noise power, and $\bar{\Psi}_y $ is the point process of satellites visible at $ y $. The transmit or association satellite is assumed to direct its beam toward its intended receiver. Namely we let $ G_{X_y;y} = g g_r = g $ and $ G_{X_j;y} =g_r=1$ for all $  X_j \neq X_y $. We use subscripts to distinguish satellites' locations and small-scale fading associated with them. The constant $ \tau $ is the SINR threshold. }

	\section{Statistics of Orbits and Satellites}\label{S:properties}
	This section provides statistical properties of the proposed satellite Cox point process that are essential to the analysis of the downlink communications from satellites to users. 
	\subsection{Isotropy}

\begin{theorem}\label{lemma:2}
	$\mathcal O$ and $\Psi$ are isotropic, namely invariant by all rotations. 
\end{theorem}

\begin{IEEEproof}
Below, the reference basis of $\mathbb{R}^3$ is denoted by $(e_x,e_y,e_z)$ and 
the unit sphere of center $o=(0,0,0)$ in $\mathbb{R}^3$ is denoted by $\bS$. 
Let $ U $ be a uniformly distributed random point on $\bS$. 
For each ${U}$, there is a unique directed orbit $\cO (U)\subset \bS$, which is the orbit whose normal vector is $\overrightarrow{oU}$
with a direction that is the trigonometric (counterclockwise) direction with respect to (w.r.t.) this vector (namely seen from point $U$).
Note that $\cO (U)$ is a factor of ${U}$; namely, for all $\bR^3$ rotations $R$ of center $o$, 
the orbit $\cO (R(U))$ coincides with the orbit $R(\cO (U))$.
Since the law of $U$ is isotropic on $\bS$ (namely left invariant by all $R$), it follows from the
relation $\cO (R(U))=R(\cO (U))$ that the law of $\cO (U) $ is also isotropic.

It is well known \cite{muller1959note} that the uniform random vector ${U}$ can be represented as 
\begin{equation}
	{U} = (\sqrt{1-V^2}\cos(\Theta), \sqrt{1-V^2}\sin(\Theta), V), \label{eq:1}
\end{equation} 
in the basis $(e_x,e_y,e_z)$ where  
$V\sim\text{Uniform}(-1,1)$,  $\theta\sim\text{Uniform}(0,2\pi),$ and $V\independent \Theta.$  

For the directed orbit $\cO (U)$, we define 
\begin{itemize}
	\item The longitude angle $\theta\in [0,2\pi)$ to be the angle $\angle{e_x}\overrightarrow{oA}$, where
	$A$ is the ascending point of the orbit on the $xy$-plane; 
	\item The inclination angle $\phi\in [0,\pi)$ to be the angle $\angle{}e_z\overrightarrow{oU}$.
\end{itemize}
It follows from  Eq \eqref{eq:1} that we can represent the longitude and inclination as follows: 
\begin{align}
		\theta& = \Theta+\pi/2 \mod 2\pi,\\
	\phi & = \arccos(V).
\end{align}
Since $V$ and $\Theta$ are independent random variables, we have that $\phi$  and $\theta$ are independent. In other words, the longitude and inclination of the orbit are independent. 
Furthermore, since $\Theta\sim\text{Uniform}[0,2\pi)$, we also have $\theta \sim \text{Uniform}[0,2\pi)$.
Then, based on the fact that $V\sim\text{Uniform}[-1,1)$, we arrive at 
\begin{align}
	{\mathbb{P}}(\phi<x) = {\mathbb{P}}(V<\cos(x)) = \frac{1-\cos(x)}{2}, \quad\quad  0 \leq  x <\pi.\nnb
\end{align}
Using the above CDF, we get the PDF of $\phi$ as follows: 
\begin{equation}
	f_{\phi}(x) =\begin{cases}
		\frac{\sin(x)}{2}, &\text{for }0\leq x< \pi, \\
		0 & \text{otherwise}.
	\end{cases}
\end{equation}
The isotropic directed orbit Poisson point process can hence be represented as a Poisson point process of density
\begin{equation}
	\Lambda(\phi, \theta) =\frac{ \lambda}{4\pi}\sin(\phi),
\end{equation}
on the rectangle set $\mathcal{R} = [0,\pi)\times [0,2\pi)$.
Here, $\lambda$ corresponds to the mean number of directed orbits. 

Furthermore, the directed orbit with angles $(\theta,\phi)$ and that with angles $(\theta+\pi, 
\phi+\pi/2 \mod \pi)$ reduce to the same orbit when forgetting the orbit direction.
Therefore, for an undirected isotropic orbit, its longitude angle
$\widetilde \theta$ is defined as the angle that the orbital plane makes with the reference plane in $[0,\pi)$ and it is uniformly distributed in this interval.

Based on the same principle, the isotropic undirected orbit Poisson point process can hence be represented as a Poisson point process of density
\begin{equation}
	\widetilde{\Lambda}(\widetilde \phi, \widetilde \theta) = \frac{\widetilde \lambda}{2\pi}\sin(\widetilde \phi) , \label{11}
\end{equation}
on the rectangle set $\widetilde{\mathcal{R}} = [0,\pi]\times [0,\pi]$. Here, $\widetilde \lambda$ is the mean number of undirected orbits. 

Since the density of the orbit process is given by Eq. \eqref{11}, the proposed orbit process is isotropic. Furthermore, conditionally on each orbit, the satellite point process on each orbit is isotropic. As a result, the Satellite Cox point process is isotropic. 
\end{IEEEproof}
\begin{remark}
\cc{Note that this isotropy property also appears in recent work \cite{9079921,9177073,9218989,9497773,9678973,9861782} where binomial or Poisson point processes were used to model the distribution of LEO or MEO satellites. Those isotropic models are capable of effectively depicting the local geometry of non isotropic satellite deployment scenarios, by using the mean parameters obtained locally from such scenarios at a given latitude. Similarly, our developed framework is also capable of reproducing the local geometry of satellite distribution and analyzing the downlink communications therein, by finding the mean number of orbits and the mean number of satellites per orbit which are based on a real-world scenario at a given latitude and then applying them in the analysis. See Section \ref{S:6}.}
\end{remark}
	
	\begin{lemma}\label{Lemma:no_satellite}
		The average number of all satellites is $ \lambda\mu. $
	\end{lemma}
	\begin{IEEEproof}
		The average number of satellites is given by 
		\begin{align}
			\bE[\Psi] &\ea \bE\left[\sum_{l(\theta_i,\phi_{i})\in \cO}\sum_{X_j\in \psi_i} 1 \right]\nnb\\
			&\eb\bE\left[\sum_{l(\theta_i,\phi_{i})\in \cO}\bE\left[\left.\sum_{X_j \in \psi_i}1 \right| \cO\right]\right]\nnb\\
			&\ec\bE\left[\sum_{l(\theta_i,\phi_{i})\in \cO}\frac{\mu}{2\pi } \int_{0}^{2\pi}\diff \omega   \right]\nnb\\
			&\ed  {\mu}\int_{0}^{\pi}\int_0^{{\pi}} \frac{\lambda\sin(\phi)}{2\pi}  \diff \theta \diff \phi  	= \lambda \mu,
		\end{align}
		where (a) follows from the definition of the number of points on all orbits. By conditioning on $ \cO, $ we have (b). Since the satellite point process on the orbit $ l(\theta_i,\varphi_i) $ is created by the Poisson point process $ \phi_i $ of intensity $ \mu/(2\pi) $ on $ \cI $, we can use Campbell's mean value theorem \cite{baccelli2010stochastic} to get (c). We obtain (d) from Campbell's mean value theorem on the Poisson point process $ \Xi $.  
	\end{IEEEproof}

\begin{remark}
	In this paper, we consider a typical user at $n=(0,0,r_e).$ Since the satellite Cox point process is isotropic and the users are independent of the satellites, the typical user at the above location can represent all the users in the network. In other words, the probability law of a satellite network seen from the typical user is the same as that of the satellite network seen from any users at any locations on Earth. Therefore, the average number of satellites visible from the typical user is the mean number of satellites visible from any users on the Earth. Moreover the statistics of the SINR or the interference seen by the typical user represent the distributions of the SINR or interference of any user in the network. 
\end{remark}
Below, we evaluate the average number of satellites visible from the typical user. 
First, let us define a spherical cap as follows: 
	\begin{equation}
	C_{d} = \{(x,y,z)\in\bS_{r_s}| \sqrt{x^2+y^2+(z-r_e)^2}\leq d \},\label{eq:scap}
\end{equation}
where $ d $ is a positive constant. This spherical cap is the set of points on $ \bS_{r_s} $ the sphere of radius $r_s$ whose distance to the typical user is less than or equal to $ d $. Fig. \ref{fig:figure1} shows the spherical caps $ C_{d} $ and $ C_{\overline{d}} $, where $ r_s-r_e<d<\overline{d} $ and $ \overline{d}=\sqrt{r_s^2-r_e^2}$, the maximum distance to a visible satellite.  


\begin{proposition}\label{cor:1}
From any user on the Earth, the average number of visible satellites is given by  	
\begin{align}
\frac{\lambda\mu}{\pi}\int_0^{\bar{\varphi}} \cos(\varphi)\arcsin\left(\sqrt{1-\cos^2(\overline{\varphi})\sec^2(\varphi)}\right)\diff \varphi\label{eq:proposition},
\end{align}
where $\overline{\varphi}=\arccos(r_e/r_s).$
\end{proposition}
\begin{IEEEproof}
	Consider the spherical cap $		C_{\overline{d}} $. Let $X_{j,i}$ be the location of the $j$-th satellite on the $i$-th orbit. Then, the average number of visible satellites from the typical user is given by 
			\begin{align}
	&\bE\left[\sum_{l(\theta_i,\phi_{i})\in \cO}\sum_{X_{j,i}\in l(\theta_i,\phi_i)} \ind_{X_{j,i} \in  C_{\overline{d}}}\right]\nnb\\
		&=\bE\left[\sum_{l(\theta_i,\phi_{i})\in \cO}\bE\left[\left.\sum_{X_{j,i} \in \psi_i}\ind_{X_{j,i}\in  C_{\overline{d}}}\right| O\right]\right]\nnb\\
		&\ea\bE\left[\sum_{l(\theta_i,\phi_{i}) \cap C_{\overline{d}}} \frac{\mu}{{2\pi }} \frac{\text{length}(l(\theta_{i}, \phi_{i})\cap C_{\overline{d}})}{r_s}\right]\nnb\\
		&\eb\frac{\mu}{\pi}\bE\left[\sum_{l(\theta_i,\phi_{i}) \cap C_{\overline{d}}\neq \emptyset} \arcsin\left(\sqrt{1-\cos^2(\overline{\varphi})\csc^2(\phi_i)}\right)\right]\nnb.
	\end{align}
To obtain (a), we use the fact that the mean number of satellites on the arc  $ C_{\overline{d}}\cap l(\theta,\phi)$ is given by the product of the angle of the arc and of the intensity  $ \mu/(2\pi ). $  To get (b), we use the result in Appendix \ref{length of an arc}, which gives the arc length of the intersection of an orbit and a spherical cap.  

Then, employing Campbell's averaging formula, the average number of visible satellites is given by  
\begin{align}
	&\frac{\mu\lambda}{2\pi}\int_{\pi/2-\bar{\varphi}}^{\pi/2+\overline{\varphi}} {\sin(\phi)}\arcsin\!\left(\sqrt{1-\cos^2(\overline{\varphi})\csc^2(\phi)}\right)\diff \phi\nnb\\
	&=\frac{\lambda\mu}{\pi}\int_0^{\bar{\varphi}} \cos(\varphi)\arcsin\left(\sqrt{1-\cos^2(\overline{\varphi})\sec^2(\varphi)}\right)\diff \varphi.\nnb
\end{align}
To get the first expression, we use the fact that an orbit $ l(\theta_i,\phi_{i}) $ meets $ C_{\overline{d}} $ if and only if its inclination satisfies $ |\pi/2-\phi_i|\leq \overline{\varphi}=\arccos(r_e/r_s). $ Then, we use the change of variables $\pi/2-\phi = \varphi $ to obtain the final result. 
 \end{IEEEproof}
In Eq. \eqref{eq:proposition}, the integration gives a constant. For instance, in a case of $r_a=525$,  the integration gives $0.038$ and therefore, the mean number of visible satellites is given by $0.038\lambda\mu$. In a case of $ r_a=1100 $  km, the integral formula gives $0.074$ and there are about $0.074\lambda\mu$ visible satellites on average.

\section{No-Satellite Probability}
\subsection{No-Satellite Probability}

We leverage Theorem \ref{lemma:2} to obtain performance metrics of network users by deriving those of the typical user.  In this section, we derive (i) the no-satellite probability $ \bP(\text{no-satellite}) $ and (ii) the distance distribution from the typical user to the association satellite, namely nearest visible satellite, $ \bP(D>d). $

		\begin{theorem}\label{T:no-satellite}
			The no-satellite probability of the typical user is 
						\begin{align}
				e^{-{\lambda}\int_{0}^{\overline{\varphi}}\cos(\varphi)\left(1- e^{-\frac{\mu}{\pi}\arcsin(\sqrt{1-r_e^2\sec^2(\varphi)/r_s^2})}\right)\diff \varphi }.\label{eq:Theorem2}
			\end{align}
		\end{theorem}
		\begin{IEEEproof}
			Let $ D $ be the distance from the typical user to its nearest satellite. Then, the typical user observes no satellite if and only if the distance from the typical user to its nearest satellite $ D $  is greater than $ \overline{d} $. The no-satellite probability is given by 
			\begin{align}
				&\bP(D>\overline{d})\nnb\\
				& = \bP( \|X_i-n\|\geq \sqrt{r_s^2-r_e^2}, \forall  X_i\in \Psi)\nnb\\ 
				&=\bP\left(\bigcap_{(\theta_i, \phi_i)\in \Xi}\left(\bigcap_{X_{j,i} \in \psi_{i}}\|X_{j,i}-n\|\geq \overline{d}\right)\right)\nnb\\
				&=\bE\left[\prod_{(\theta_i, \phi_i)\in \Xi}\bE\left[\left.\prod_{X_{j,i}\in \psi_{i}}\ind_{\|X_{j,i}-n\|>\overline{d}}\right| \Xi\right]\right],\label{eq:outage-1}
			\end{align}
		where we use the fact that, conditionally on orbits, the Poisson point processes on orbits $ \{\psi_i\}_{i} $ are independent.  
		\par To evaluate Eq. \eqref{eq:outage-1}, we consider the spherical cap $ C_{\overline{d}} $. Then, we use the fact that the event that all satellites on orbit $ l(\theta_i,\phi_i) $ are located at distances greater than $ \overline{d} $ is equivalent to the event that the orbit $ l(\theta_i,\phi_i) $ contains no satellite on $ C_{\overline{d}} $. By using the density of the satellite Poisson point process $ \psi_i $, we get 
	\begin{align}
		\bE\left[\left.\prod_{X_{j,i} \in \psi_{i}}\ind_{\|X_{j,i}-n\|>\overline{d}}\right| \Xi \right]= e^{-\frac{\mu}{2\pi} {\frac{1}{r_s}}\nu(C_{\overline{d}}\cap l(\theta_i,\phi_i) )}, \nnb
	\end{align}
where $ \nu(C_{\overline{d}}\cap l(\theta_i,\phi_i) ) $ is the length of the arc $ C_{\overline{d}} \cap  l(\theta_i,\phi_i) $ given by 
\begin{equation*}
	\nu(C_{\overline{d}}\cap l(\theta_i,\phi_i) )  = 2r_s\arcsin(\sqrt{1-\cos^2(\overline{\varphi})\csc^2(\phi_{i})}),
\end{equation*}
for $ \pi/2-\overline{\varphi}<\phi_{i}<\pi/2+\overline{\varphi} $. Therefore, we have 
\begin{align}
	&\bP(D>\overline{d})\nnb\\
	&= \bE\left[\prod_{(\theta_i, \phi_i)\in \Xi}^{|\phi_i-\pi/2|<\overline{\varphi}}e^{-\mu\pi^{-1}\arcsin(\sqrt{1-\cos^2(\overline{\varphi})\csc^2(\varphi_{i})})}\right]\nnb\\
	&=e^{-\frac{\lambda}{2}\int_{\pi/2-\overline{\varphi}}^{\pi/2+\overline{\varphi}} \sin(\phi)\left(1- e^{-\frac{\mu}{\pi}\arcsin(\sqrt{1-\cos^2(\overline{\varphi})\csc^2(\phi)})}\right)\diff \phi }\nnb\\
	&=e^{-{\lambda}\int_{0}^{\overline{\varphi}} \cos(\varphi)\left(1- e^{-\frac{\mu}{\pi}\arcsin(\sqrt{1-\cos^2(\overline{\varphi})\sec^2(\varphi)})}\right)\diff \varphi }\nnb,
\end{align}
where we used the probability generating functional of the Poisson point process of density $ \lambda\sin(\phi)/2\pi $ on the rectangle $\cR$ and then the change of variables $\varphi = \pi/2-\phi$. 
		\end{IEEEproof}	
The above no-satellite probability gives the probability that an arbitrarily located network user on the Earth is not able to find any satellite at a given time. The derived probability expression is a function of the network geometric parameters $\lambda,\mu$, and $r_s$. 
\par
\begin{figure}
	\centering
	\includegraphics[width=.9\linewidth]{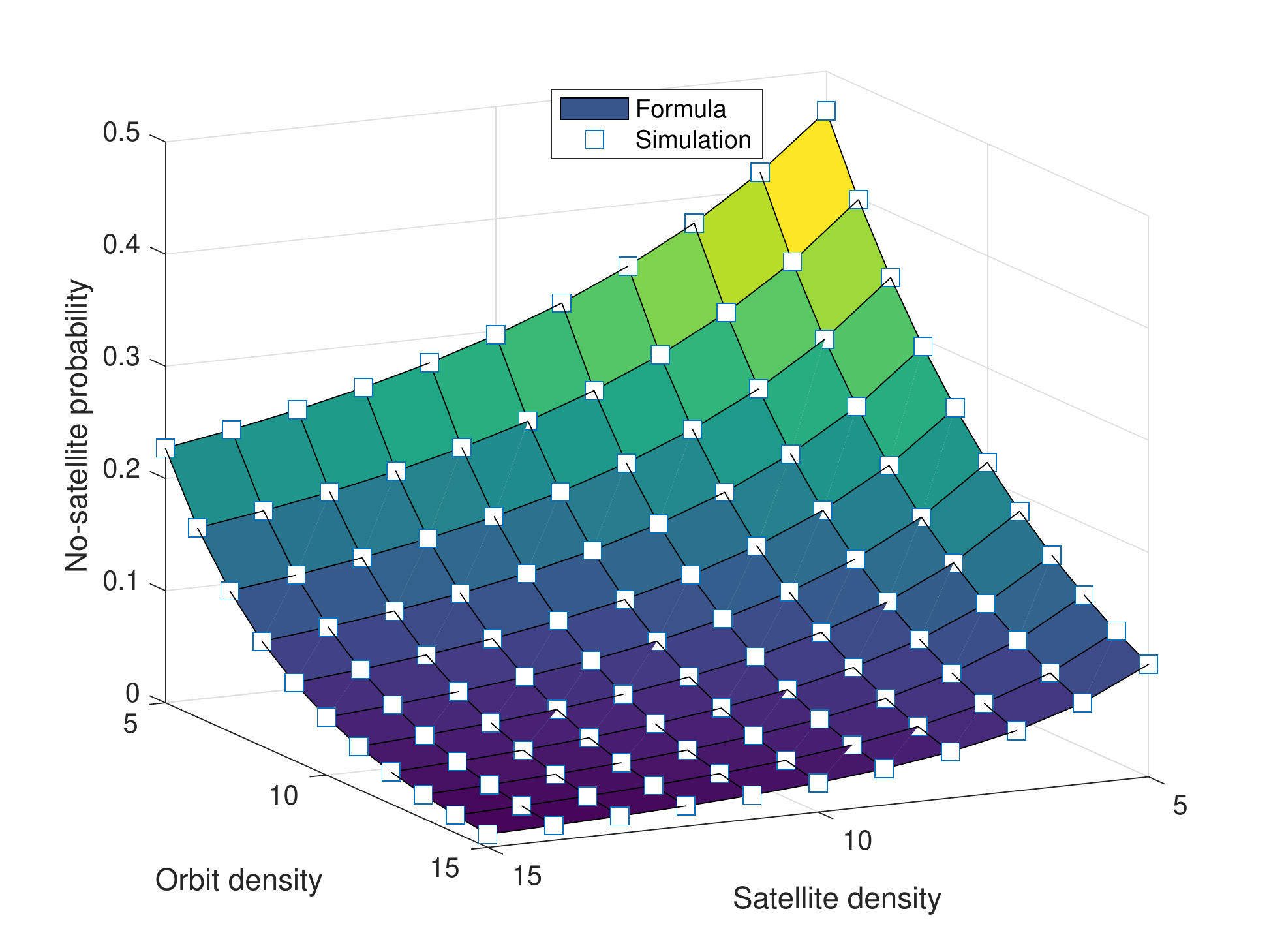}
	\caption{The no-satellite probability. The satellite altitude is $550$ km. Simulation results validates the derived formula.}
	\label{fig:nosatelliteprobability550km}
\end{figure}

\begin{figure}
	\centering
	\includegraphics[width=.9\linewidth]{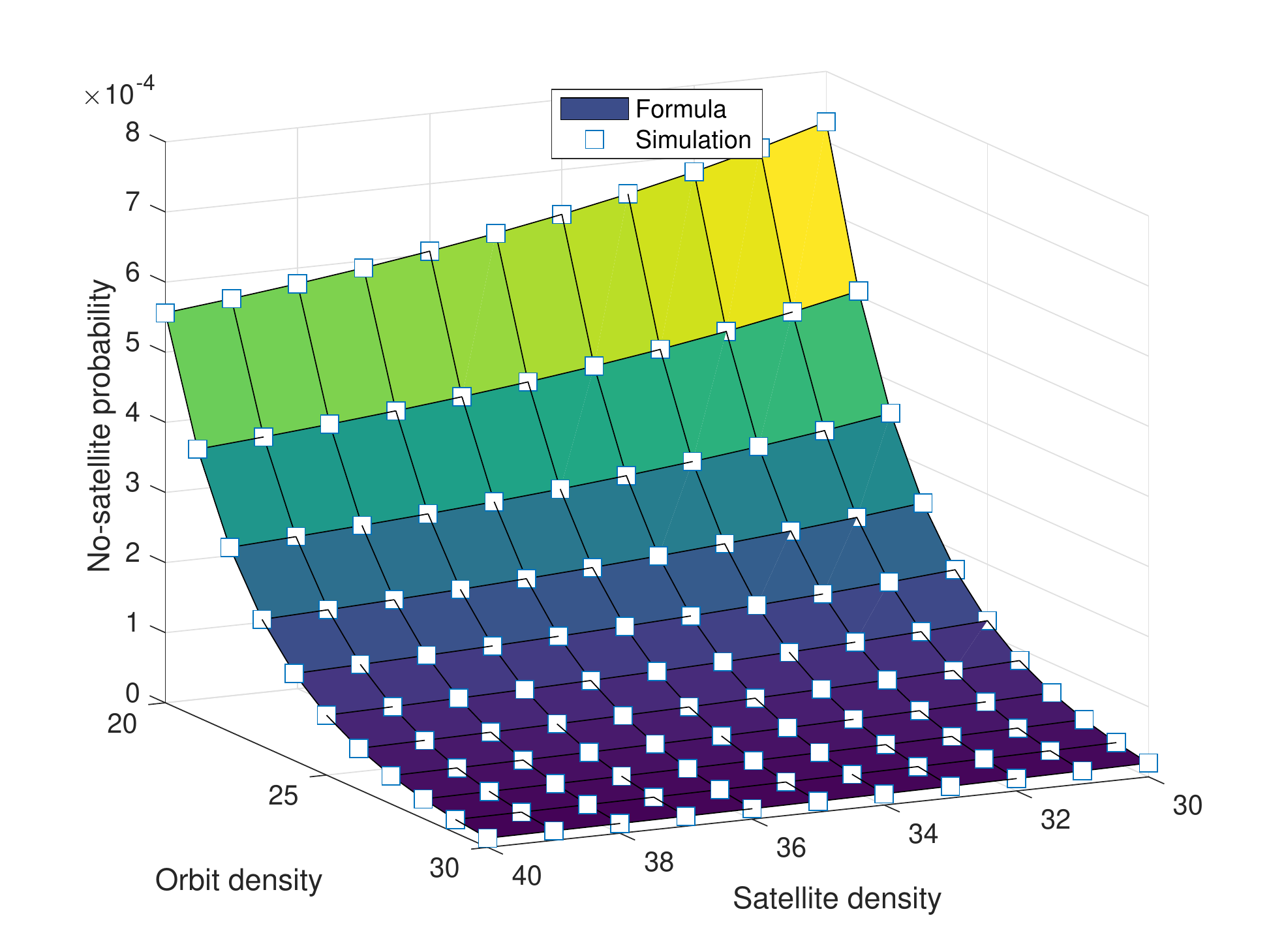}
	\caption{The no-satellite probability. The satellite altitude is $550$ km. Simulation results validates the derived formula.}
	\label{fig:nosatelliteprobability550kmdense}
\end{figure}

\begin{figure}
	\centering
	\includegraphics[width=.9\linewidth]{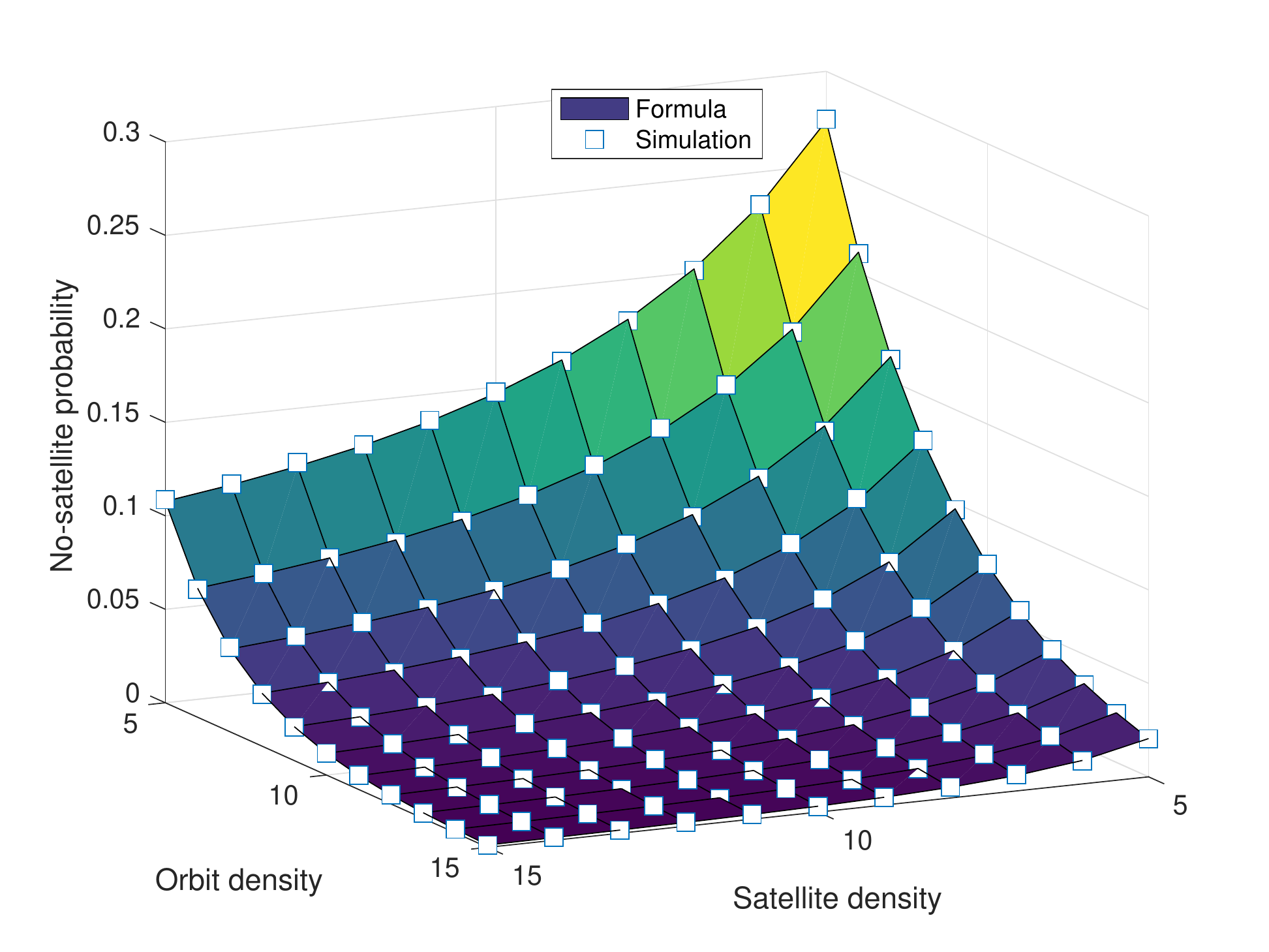}
	\caption{The no-satellite probability. The satellite altitude is $1100$ km. Simulation results validates the derived formula.}
	\label{fig:nosatelliteprobability}
\end{figure}

 Since $ r_s, \lambda $ and $ \mu $ are system parameters, large-scale impact of modifying geometry of the satellite network, such as increasing the number of orbits or decreasing the satellite density, can easily be evaluated. For instance, we observe that the no-satellite probability expression of Theorem \ref{T:no-satellite}  is {exponential} in the number of orbits. Figs. \ref{fig:nosatelliteprobability550km} and \ref{fig:nosatelliteprobability550kmdense} show the no-satellite probability when the satellite altitude is $550$ km. The simulation results are produced by Monte Carlo simulation with the sample size $N=10^6$. For each simulation instance, we create orbits and their satellites based on the definition of the Cox point process. Then, we count the event that there is no visible satellite at the typical user and combine such events to get the probability. We confirm that the simulation results validate the accuracy of the derived formula in Theorem \ref{T:no-satellite}. It is important to note that for a reasonably many satellites on orbits, e.g., $\lambda\geq 25$ and $\mu\geq 25$, the no-satellite probability is less than $0.001$. Fig.  \ref{fig:nosatelliteprobability} also displays the no-satellite probability when the satellite altitude is $1100$ km and $\lambda\in(5,15)$ and $\mu\in(5,15).$ Again, the simulation results validate the accuracy of the  derived formula in Theorem \ref{T:no-satellite}. 

\par For the moment, suppose $ \mu $ is very high. In this case, each orbit has a very large number of satellites and thus the no-satellite probability is strongly dictated by the distribution of orbits. From the above formula, with $\mu\gg1$, we have $ e^{-2r_s\mu} \approxeq 0 $ and thus the above derived no-satellite probability is asymptotically approximated as $		\exp\left(-{\lambda}\sin(\overline{\varphi})\right).$ For instance, if $ r_s = 7000 $ km, we have $ \exp(-0.4\lambda)$. If $ \lambda = 52 $, the no-satellite probability is approximately $10^{-9}. $


\subsection{Distribution of the Distance to the Nearest Satellite}
 Let $ D $ be the distance from the typical user to its nearest satellite. Note that when the nearest satellite is visible, we have $ r_s-r_e\leq D\leq r_s^2-r_e^2$. We set $ D= \infty $ if there is no visible satellite from the typical user.  
		\begin{theorem} 
$\bP(D>d)$ the CCDF of the distance to the nearest satellite is given by $1$ for $d<r_s-r_e$. Then,  for 		$ r_s-r_e \leq d <\sqrt{r_s^2-r_e^2}$, 
			\begin{align}
	&e^{-{\lambda}\int_{0}^{\xi (d)} \cos(\varphi)\left(1- e^{\-\frac{\mu}{\pi}\arcsin\left(\sqrt{1-\cos^2(\xi(d))\sec^2(\varphi)}\right)}\right)\diff \varphi }.\nnb
\end{align}
where 	$\xi(d) = \arccos\left(({r_s^2+r_e^2-d^2})/({2r_er_s})\right)$. Then, for $d\geq \sqrt{r_s^2-r_e^2},$ it is equal to the no-satellite probability. 
 		\end{theorem}
	\begin{IEEEproof}
The distance to the nearest visible satellite is greater than $ d $ if and only if all the satellites are at distances greater than $ d $. Since the orbit radius is $r_s$, the CCDF is $1$ for $d<r_s-r_e.$
 
For $r_s-r_e\leq  d<\sqrt{r_s^2-r_e^2}, $ we exploit Appendix \ref{length of an arc} and the proof of Theorem \ref{T:no-satellite} to obtain 
		\begin{align}
		&\bP(D>d) \nnb\\
		&=e^{-{\lambda}\int_{\pi/2-\xi(d)}^{\pi/2+\xi(d)}\frac{\sin(\phi)}{2}\left(1- e^{-\frac{\mu}{\pi}\sin^{-1}\left(\sqrt{1-\cos^2(\xi(d))\csc^2(\phi)}\right)}\right)\diff \phi}\nnb\\
		&=e^{-{\lambda}\int_{0}^{\xi}\cos(\varphi)\left(1- e^{-\frac{\mu}{\pi}\sin^{-1}\left(\sqrt{1-\cos^2(\xi(d))\sec^2(\varphi)}\right)}\right)\diff \varphi},\nnb
	\end{align}
where we use the fact that only the orbits having inclinations between $\pi/2-\xi(d)$ and $\pi/2+\xi(d)$ will meet the spherical cap $C_{d}$. For $r_s-r_e\leq d<\sqrt{r_s^2-r_e^2},$ we have  
\begin{equation*}
\xi(d) = \arccos\left(({r_s^2+r_e^2-d^2})/({2r_s r_e})\right).
\end{equation*}
Note that $d$ denotes the distance to the nearest satellite and it is the value that $D$ will take. 

For $d\geq \sqrt{r_s^2-r_e^2},$ the CCDF evaluated at such $d$ is the probability that there is no visible satellite, namely the no-satellite probability of Theorem \ref{T:no-satellite}. 
\end{IEEEproof}

\section{Coverage Probability}
This section evaluates the coverage probability of the typical user. We start with the Rayleigh fading case. Table \ref{Table:1} summarizes the system-level simulation parameters used for the SINR compuation.

\begin{theorem}\label{mainT}
	In the interference-limited regime with $ m=1 $, namely in Rayleigh fading environment, the coverage probability of the typical user  is given by  Eq. \eqref{eq:theorem:4a}.
\end{theorem}

\begin{figure*}
\begin{align}
&\int_{r_s-r_e}^{\sqrt{r_s^2-r_e^2}}\frac{\lambda \mu z }{ \pi r_sr_e}\exp\left(-{\lambda}\int_{\xi(z)}^{\bar{\varphi}}\left(1- e^{-\frac{\mu}{\pi}\int_{0}^{\omega_{2}(\nu)}\left(1-\cL_H(\frac{\tau z^\alpha}{g\overline{K_{\nu,\omega}}^{\alpha}})\right)\diff \omega}\right)\cos(\nu)\diff \nu\right)\nnb\\ 
		&\hspace{17mm}\times\exp\left(-{\lambda}\int_{0}^{\xi(z)}\left(1- e^{-\frac{\mu}{\pi}\arcsin(\sqrt{1-\cos^2(\xi(z))\sec^2(\nu)})-\frac{\mu}{\pi}\int_{\omega_{1}(\nu,z)}^{\omega_{2}(\nu)}\left(1-\cL_H(\frac{\tau z^\alpha}{g\overline{K_{\nu,\omega}}^{\alpha}})\right)\diff \omega}\right)\cos(\nu)\diff \nu\right)\nnb\\
		&\hspace{17mm}\times \left(\int_{0}^{{\xi(z)}}\frac{e^{-\frac{\mu}{\pi}\arcsin(\sqrt{1-\cos^2(\xi(z))\sec^2(\varphi)})-\frac{\mu}{\pi}\int_{\omega_{1}(\varphi,z)}^{\omega_{2}(\varphi)}\left(1-\cL_H(\frac{\tau z^\alpha }{g\overline{K_{\varphi,\omega}}^{\alpha}})\right)\diff \omega}}{\sqrt{1-\cos^2(\xi(z))\sec^2(\varphi)}}\diff \varphi\right)  \diff z \label{eq:theorem:4a}.
	\end{align}
	\rule{1\linewidth}{0.2mm}
\end{figure*}

\begin{IEEEproof}
		Let $ X_\star$ denote the location of the satellite nearest to the typical user at $n$,
$I$ the power of the interference created by the visible satellites at distances greater than
$\|X_\star-n\|$, and $g$ the aggregate antenna gain when the transmit and receiver
antennas are aligned to provide the maximum antenna gain, namely $g=g_1+g_2$. We have	
\begin{eqnarray}
	\bP(\SINR>\tau)
	= & \bE\left[\cL_{I}\left(\frac{\tau \|X_\star-n\|^{\alpha}} {pg}\right)\right]	\label{eq:a26},
\end{eqnarray}
where $\cL_{I}(s)$ is the Laplace transform of $I$.
We used here the fact that, for $m=1,$ $H$ follows an exponential random variable with mean one independent of $I$ and $X_\star$.

The difficulty with Eq. (\ref{eq:a26}) is that the random variables $I$ and $\|X_\star-n\|$ are not independent.
In order to compute their joint distribution, we proceed in 3 main steps.

The first step consists in introducing a partition of all possibilities concerning $D$ and the orbit that contains $X_\star$.
Let $i_\star$ denote the index of the latter. We have
\begin{align}
	&  \bE\left[\cL_{I}\left(\frac{\tau \|X_\star-n\|^{\alpha}} {pg}\right)\right]	
	\label{eq:26bis}\\
	& =  \int\limits_z  
	\bE \left[\sum_i \cL_{I}\left(\frac{\tau z^{\alpha}} {pg}\right) 1_{i_\star=i} 1_{D\in [z,z+dz]}\right] 
	\nnb\\
	& \ea  \int\limits_z  \int\limits_{(\theta,\phi)}
	\bE^0_{\theta,\phi}\left[ \cL_{I}\left(\frac{\tau z^{\alpha}} {pg}\right)
	1_{l_{i_\star}=l(\theta,\phi)} 1_{D\in [z,z+dz]}\right]\nnb\\ 
	& \hspace{4cm}\Lambda(\diff\theta,\diff \phi)\nnb\\
	& \eb  \lambda \int\limits_z  \int\limits_{\phi}
	\bE^0_{\theta,\phi}\left[ \cL_{I}\left(\frac{\tau z^{\alpha}} {pg}\right) 1_{\phi_{i_\star}=\phi} 1_{D\in [z,z+dz]}\right]
	\sin(\phi) \diff \phi,\nnb
\end{align}
where $\bE^0_{\theta,\phi}$ denotes the Palm probability of $\Xi$.
In (a), we used Campbell's formula.
In (b), we used that fact that the function of interest are invariant by a change of $\theta$.
Note that the law of $\Xi$ under the Palm probability in question is the law of
$\Xi+\delta_{(\theta,\phi)}$, with $\Xi$ distributed as above. Hence
\begin{eqnarray}
	& & \hspace{-1.5cm}
	\bE^0_{\theta,\phi}\left[ \cL_{I}\left(\frac{\tau z^{\alpha}} {pg}\right) 1_{\phi_{i_\star}=\phi} 1_{D\in [z,z+dz]}\right] \nnb
	\\
	& = & \bE\left[ \cL_{\widehat I}\left(\frac{\tau z^{\alpha}} {pg}\right)
	1_{\phi_{\widehat i_\star}=\phi} 1_{\widehat D\in [z,z+dz]}\right],\label{eq:tt1}
\end{eqnarray}
with $\widehat I$, $\widehat i_\star$ and $\widehat D$ for $\Xi+\delta_{(\theta,\phi)}$ in place of $\Xi$.

Let $\mathcal F$ be the sigma-algebra generated by $\Xi$. We have
\begin{eqnarray}
	& & \hspace{-.6cm} \bE\left[ \cL_{\widehat I}\left(\frac{\tau z^{\alpha}} {pg}\right)
	1_{\phi_{\widehat i_\star}=\phi} 1_{\widehat D\in [z,z+dz]}\right]\nnb\\
	& = &
	\bE \left[ \bE\left[ \cL_{\widehat I}\left(\frac{\tau z^{\alpha}} {pg}\right)
	1_{\phi_{\widehat i_\star}=\phi} 1_{\widehat D\in [z,z+dz]}
	\mid {\mathcal F}\right] \right]\label{eq:tt2}
\end{eqnarray}
and, by first principles,
\begin{eqnarray}
	& & \hspace{-1cm} \bE\left[ \cL_{\widehat I}\left(\frac{\tau z^{\alpha}} {pg}\right)
	1_{\phi_{\widehat i_\star}=\phi} 1_{\widehat D\in [z,z+dz]}\mid {\mathcal F}\right]	\label{eq:bfp}\\
	& = & 
	\bE\left[ \cL_{\widehat I}\left(\frac{\tau z^{\alpha}} {pg}\right) \mid 
	\phi_{\widehat i_\star}=\phi,  \widehat D\in [z,z+dz], {\mathcal F}\right]\nnb\\
	& & \hspace{2cm}\bP \left[\phi_{\widehat i_\star}=\phi,  \widehat D\in [z,z+dz] \mid {\mathcal F}\right].\nnb
\end{eqnarray}

The next steps consist in computing the two terms in the last expression.
We start with the conditional Laplace transform of the interference $\widehat I$. As we now show,
conditionally on $\mathcal F$ and the event $\widehat{\mathcal{G}}(z)$ that
$\phi_{\widehat i_\star}=\phi$ and $ \widehat D\in [z,z+dz]$, the interference created
by the satellites on different orbits are conditionally independent and there is a closed form
for the Laplace transform of the interference on each orbit.
To derive this closed form, we use 
\begin{align}
	\xi(z)&=\arccos\left({(r_s^2+r_e^2-z^2)}/{(2 r_sr_e)}\right).\label{25}
\end{align}
The angle $\xi(z)$ is that between the north direction
and any point on the rim of the spherical cap $C_z$ for $r_s-r_e<z<\bar d =\sqrt{r_s^2-r_e^2}$. Then we need the following 
\begin{align}
	K_{i,\omega}&=\sqrt{r_s^2-2r_s r_e\sin(\omega)\sin({\phi}_i)+r_e^2},\\
	K_{\omega}&=\sqrt{r_s^2-2r_s r_e\sin(\omega)\sin({\phi})+r_e^2},\\
	\kappa_{i,1}&={\pi}/{2}-\arcsin(\sqrt{1-(r_e/r_s)^2\csc^2(\phi_i)}),\label{a23}\\
	\kappa_{1}&={\pi}/{2}-\arcsin(\sqrt{1-(r_e/r_s)^2\csc^2(\phi)}),\label{a23b}\\
	\kappa_{i,2}(z)& = {\pi}/{2}-\arcsin(\sqrt{1-{\cos^2(\xi(z))}{\csc^2(\phi_i)}}),\label{a24}\\
	\kappa_{2}(z)& = {\pi}/{2}-\arcsin(\sqrt{1-{\cos^2(\xi(z))}{\csc^2(\phi)}}).\label{a24b}
\end{align}
The function $K_{i,\omega}$ gives the distance between the typical user and the point of orbit $l(\theta_i,\phi_i)$
with orbital angle $\omega$.
The function $\kappa_{i,2}(z)$ gives the angle (in the orbital plane of orbit $i$, i.e., $l(\theta_i,\phi_i)$)
between the apex of the orbit and the point of the orbit where the distance to $n$ is $z$. 
The parameter $\kappa_{i,1}$ is the angle (in the orbital plane of orbit $i$) between the apex of the orbit
and the point of the orbit beyond which a satellite is no more visible from $n$.

Let ${\mathcal E}(z)= {\mathcal F} \cap \widehat {\mathcal G}(z)$. For short we use the notation $\mathcal E$
for ${\mathcal E}(z)$.
Let $N$ denote the number of points of $\xi$ and let $(\theta_i,\phi_i),\ i=1,\ldots,N$, denote their coordinates.
Let $\psi_{i}$ denote the linear Poisson point process on $l(\theta_{i},\phi_{i})$.
Note that these random variables are $\mathcal F$-measurable.
Using the conditional independence alluded to above, and the independence of the number of points of the linear Poisson
point processes on disjoint subsets of an orbit, we get 
\begin{align}
	\cL_{\widehat I|{\mathcal{E}}}(s)=& \bE_{\psi}\left[\prod_{X_{j} \in \psi}^{\overline d >K_{j}>z}
	\cL_H(spK_{j}^{-\alpha}) \right] \nnb\\
	&\prod_{i=1,N}^{|\phi_{i}-\pi/2|<\overline{\varphi}}
	\bE_{\psi_i}\left[\prod_{X_{j,i} \in \psi_{i}}^{\overline d >K_{i,j}>z}
	\cL_H(spK_{i,j}^{-\alpha}) \mid \mathcal{E}\right].\nnb
\end{align}
Then, we have 
\begin{align}
	\cL_{\widehat I|{\mathcal{E}}}(s)
	&= \exp\left(-\frac{\mu}{\pi}\int_{\kappa_{1}}^{\kappa_{2}(z)}
	\left(1 - \cL_H(sp K_{\omega}^{-\alpha})\right) \diff \omega\right)\nnb\\
	& \hspace{-5mm}\prod_{i=1,N}^{|\phi_{i}-\pi/2|\le \xi(z)}
	\exp\left(-\frac{\mu}{\pi}\int_{\kappa_{i,1}}^{\kappa_{i,2}(z)}
	\left(1 - \cL_H(sp K_{i,\omega}^{-\alpha})\right) \diff \omega\right)\nnb\\
	&\hspace{-5mm}\prod_{i=1,N}^{\xi(z)<|\phi_{i}-\pi/2|<\overline{\varphi}} \hspace{-.2cm}
	\exp\left(-\frac{\mu}{\pi}\int_{\kappa_{i,1}}^{\pi/2}
	\left( 1 - \cL_H(sp K_{i,\omega}^{-\alpha})\right) \diff \omega\right)\nnb,	
\end{align}
where $\cL_H(s)$ denotes the Laplace transform of the mean 1 exponential random variable. Using the change of variables $\omega\to \pi/2-\omega$, we get 
\begin{align}
\cL_{\widehat{I}|{\mathcal E}}(s)& =\exp\left({-\frac{\mu}{\pi}
		\int_{\omega_{1}(z)}^{\omega_{2}} 1 - \cL_H(sp\widetilde{K_{\omega}}^{-\alpha})} \diff\omega\right)\nnb \\
	&\hspace{-5mm}\prod_{i=1,N}^{|\phi_{i}-\pi/2|<{\xi(z)}}\exp\left({-\frac{\mu}{\pi}
		\int_{\omega_{i,1}(z)}^{\omega_{i,2}} 1 - \cL_H(sp\widetilde{K_{i,\omega}}^{-\alpha})} \diff \omega\right)\nnb\\
	&\hspace{-5mm}\prod_{i=1,N}^{\xi(z)<|\phi_{i}-\pi/2|<\overline{\varphi}}\exp\left({-\frac{\mu}{\pi}
		\int_{0}^{\omega_{i,2}} 1 - \cL_H(sp\widetilde{K_{i,\omega}}^{-\alpha})} \diff \omega\right),	\label{aLaplace}
\end{align} 
where 

\begin{align}
	\omega_{i,1}(z)&= \arcsin(\sqrt{1-{\cos^2(\xi(z))}{\sec^2(\phi_i)}}),	\label{26}\\
	\omega_{1}(z)&= \arcsin(\sqrt{1-{\cos^2(\xi(z))}{\sec^2(\phi)}}),	\label{26b}\\
	\omega_{i,2}&= \arcsin(\sqrt{1-(r_e/r_s)^2\sec^2(\phi_i)}),\label{26.1}\\
	\omega_{2}&= \arcsin(\sqrt{1-(r_e/r_s)^2\sec^2(\phi)}),\label{26.1b}\\
	\widetilde{K_{i,\omega}}&= \sqrt{r_s^2-2r_s r_e\cos(\omega)\sin({\phi}_i)+r_e^2}.\label{26.2}
\end{align}

Our third step consists in computing the following conditional probability
\begin{align}
	& \bP( \widehat {\mathcal G}(z)\mid {\mathcal F}) 
	=\bP(\exists j_\star\in \psi :  \|X_{j_\star}-n\| \in [z, z+\diff z], \nnb\\
	& \hspace{.3cm} \forall j\ne j_\star,  \|X_{j} -n\| > z,
	\forall i=1,N, \forall j,  \|X_{i,j} -n\| > z  |{\mathcal F}) \nnb.
\end{align}
Using now the fact that the length of the arc $C_z \cap l(\theta_i,\phi_i)$ is
$2 r_s \arcsin\left(\sqrt{1-\cos^2(\xi(z))\csc^2(\phi_i)}\right)$, we get
\begin{align}
	& \bP(\widehat{D}\in [z, z+\mathrm{d} z), \phi_{\widehat{i_\star}}=\phi \mid {\mathcal F}) 
	\label{eq:25a}\\
	&=\frac{\partial}{\partial z }\left(1-e^{-\frac{\mu}{\pi}\arcsin\left(\sqrt{1-\cos^2(\xi(z))\csc^2(\phi)}\right)}\right)
	\diff z \nnb\\
	&\hspace{6mm}\prod^{\pi/2-\xi(z)<\phi_i<\pi/2+\xi(z)}_{i=1,N}e^{-\frac{\mu}{\pi}\arcsin\left(\sqrt{1-\cos^2(\xi(z))\csc^2(\phi_i)}\right)}\nnb\\
	&= \frac{\mu z |\csc(\phi)| e^{-\frac{\mu}{\pi}\arcsin\left(\sqrt{1-\cos^2(\xi(z))\csc^2(\phi)}\right)} }{\pi r_er_s\sqrt{1-\cos^2(\xi(z))\csc^2(\phi)}}\nnb\\
	&\hspace{6mm}\prod^{\pi/2-\xi(z)<\phi_i<\pi/2+\xi(z)}_{i=1,N}e^{-\frac{\mu}{\pi}\arcsin\left(\sqrt{1-\cos^2(\xi(z))\csc^2(\phi_i)}\right)}\nnb,
\end{align}
where $\cos(\xi(z))$ is given by Eq. \eqref{25}. Here, we used the fact that 
conditionally on $ \mathcal F$, the satellite point processes on orbits are independent Poisson point processes. 

We now plug in Eq. \eqref{aLaplace} and \eqref{eq:25a} (together with \eqref{eq:tt1}, \eqref{eq:tt2}, and
\eqref{eq:bfp}) in \eqref{eq:26bis} to get \eqref{eq:theorem:4a}. This expression
is obtained by using the Laplace functional of the Poisson point process $\Xi$
and by the change of variable $\varphi= \pi/2-\phi$. It uses the functions
\begin{align*}
	\overline{K_{\varphi,\omega}} &=  \sqrt{r_s^2-2r_s r_e\cos(\omega)\cos(\varphi)+r_e^2},\\
	\omega_{1}(\varphi,z)&= \arcsin(\sqrt{1-{\cos^2(\xi(z))}{\sec^2(\varphi)}}),	\\
	\omega_{2}(\varphi)&= \arcsin(\sqrt{1-(r_e/r_s)^2\sec^2(\varphi)}).
\end{align*} 
This completes the proof. 
\end{IEEEproof}

\begin{figure*}
	\begin{align}
		&\int_{0}^{\infty}\int_{r_s-r_e}^{\sqrt{r_s^2-r_e^2}}\frac{\lambda \mu z }{ \pi r_sr_e}\exp\left(-{\lambda}\int_{\xi(z)}^{\bar{\varphi}}\left(1- e^{-\frac{\mu}{\pi}\int_{0}^{\omega_2({\varphi})}\left(1-\cL_H((2^u-1) z^\alpha /(g\overline{K_{\varphi,\omega}}^{\alpha}))\right)\diff \omega}\right)\cos(\varphi)\diff \varphi\right)\nnb\\ 
		&\hspace{20mm}\times\exp\left(-{\lambda}\int_{0}^{\xi(z)}\left(1- e^{-\frac{\mu}{\pi}\arcsin(\sqrt{1-\cos(\xi(z))^2\sec(\varphi)^2})-\frac{\mu}{\pi}\int_{\omega_1{(\varphi,z)}}^{\omega_2{(\varphi)}}\left(1-\cL_H((2^u-1) z^\alpha /(g\overline{K_{\varphi,\omega}}^{\alpha}))\right)\diff \omega}\right)\cos(\varphi)\diff \varphi\right)\nnb\\
		&\hspace{20mm}\times \left(\int_{0}^{{\xi(z)}}\frac{e^{-\frac{\mu}{\pi}\arcsin(\sqrt{1-\cos(\xi(z))^2\sec(v)^2})-\frac{\mu}{\pi}\int_{\omega_1{(v,z)}}^{\omega_2{(v)}}\left(1-\cL_H((2^u-1) z^\alpha /(g\overline{K_{v,\omega}}^{\alpha}))\right)\diff \omega}}{\sqrt{1-\cos^2(\xi(z))\sec^2(v)}}\diff v\right)  \diff z \diff u \label{eq:theorem:4.1}.
	\end{align}
	\rule{1\linewidth}{0.2mm}
\end{figure*}

In a case that thermal noise cannot be neglected, the coverage probability of the typical user can be assessed by scaling the formula in Theorem \ref{Theorem:3} with a constant.

\cc{Theorem \ref{Theorem:3} gives the coverage probability of the typical user in terms of the mean number of orbits, the mean number of satellites per orbit, the radius of orbits, and the transmit antenna gain. As our Cox-distributed satellite model exhibits rotation invariance, the coverage probability of the typical user located at $n$ statistically represent the coverage probabilities of all users in the network.  Consequently, the derived coverage probability inherently characterizes the fundamental performance of the LEO or MEO satellite network. By utilizing the formula, one can readily forecast and estimate the behavior of the coverage probability by varying the geometric parameters.}

\cc{In the following, we use the system-level parameter values in Table \ref{Table:1}, directly obtained and derived from values in \cite{38811,38821}. Specifically, we use a transmit antenna gain of 
	$g=20$ dB based on the satellite parameters provided in \cite[Section 6]{38821} and \cite[Section 6]{38811}, where the maximum transmit antenna gains are equal to $30$ dB or $24$ dB.}

\begin{table}
	\centering
	\caption{Simulation Parameters}\label{Table:1}
	\begin{tabular}{|c|c|}
		\hline
		System parameter      &  Value \\
		\hline
	Mean number of orbits 	$\lambda$ & $30$  \\
		\hline
Mean number of satellites per orbit		$\mu$  & $30$  \\
		\hline
Mean number of total satellites 				$\lambda\mu$   & $900$  \\
		\hline
Transmit antenna gain 	$g$	& $20$ dB \\
	\hline
Transmit antenna power $p$	 & $30$ dB  \\
		\hline
		Transmitter EIRP (pg)	 & $80$ dBm \\
		\hline
Satellite orbit radius 	$r_s$ &  $6950$ km\\
		\hline
Boltzmann  constant	 $k$ & $-228.6$ dBW/K/Hz \\
		\hline
Noise temperature $T$ & $290 $ K \\
		\hline 	
Receive antenna gain  $g_r$	& $0$ dB \\
\hline
System bandwith $B_w$ & $30$ MHz \\
		\hline
Carrier frequency		$f_c$ & $2 $ GHz \\
		\hline
	\end{tabular} 
\end{table}

\begin{figure}
	\centering
	\includegraphics[width=.9\linewidth]{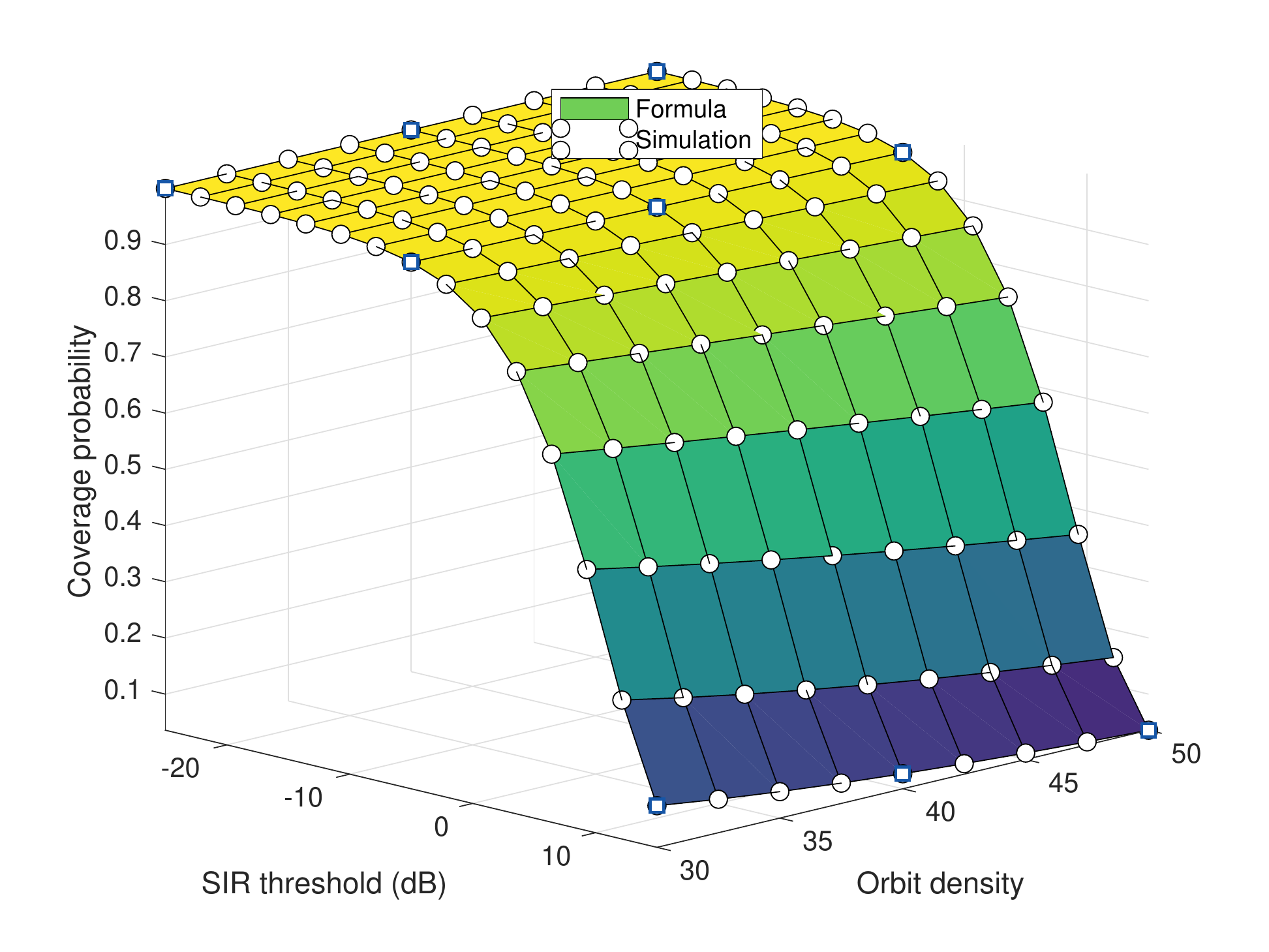}
	\caption{The coverage probability for $\lambda $. We use  $r_{s}=7050 $ km. }
	\label{fig:lambdatausir}
\end{figure}
\begin{figure}
	\centering
	\includegraphics[width=.9\linewidth]{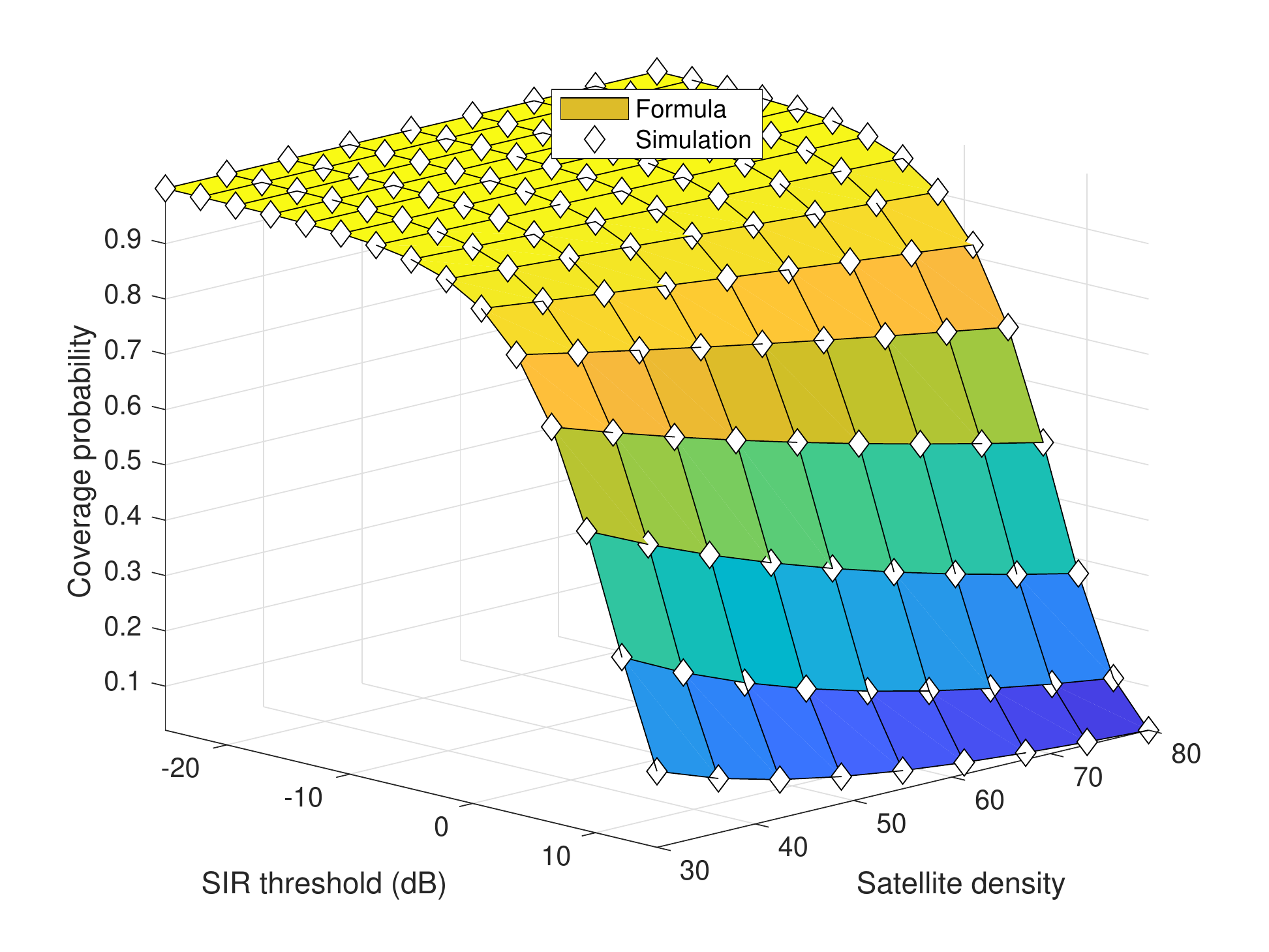}
	\caption{The coverage probability for various $\mu$. }
	\label{fig:mutausir}
\end{figure}
Fig. \ref{fig:lambdatausir} and \ref{fig:mutausir} illustrate the coverage probability of the typical user for various values of $\lambda$ and $\mu$. The simulation results validate the accuracy of the derived formula in Theorem \ref{Theorem:3}. While obtaining simulation results necessitates significant computation time due to the construction of large-scale network layout, the derived formula enables the generation of the entire coverage probability graph in much less time. This efficiency facilitates the execution of complex multivariate analyses of satellite networks, that we will see shortly. In the following, we explore the coverage behavior of the satellite network by varying various geometric parameters.


\begin{figure}
	\centering
	\includegraphics[width=.9\linewidth]{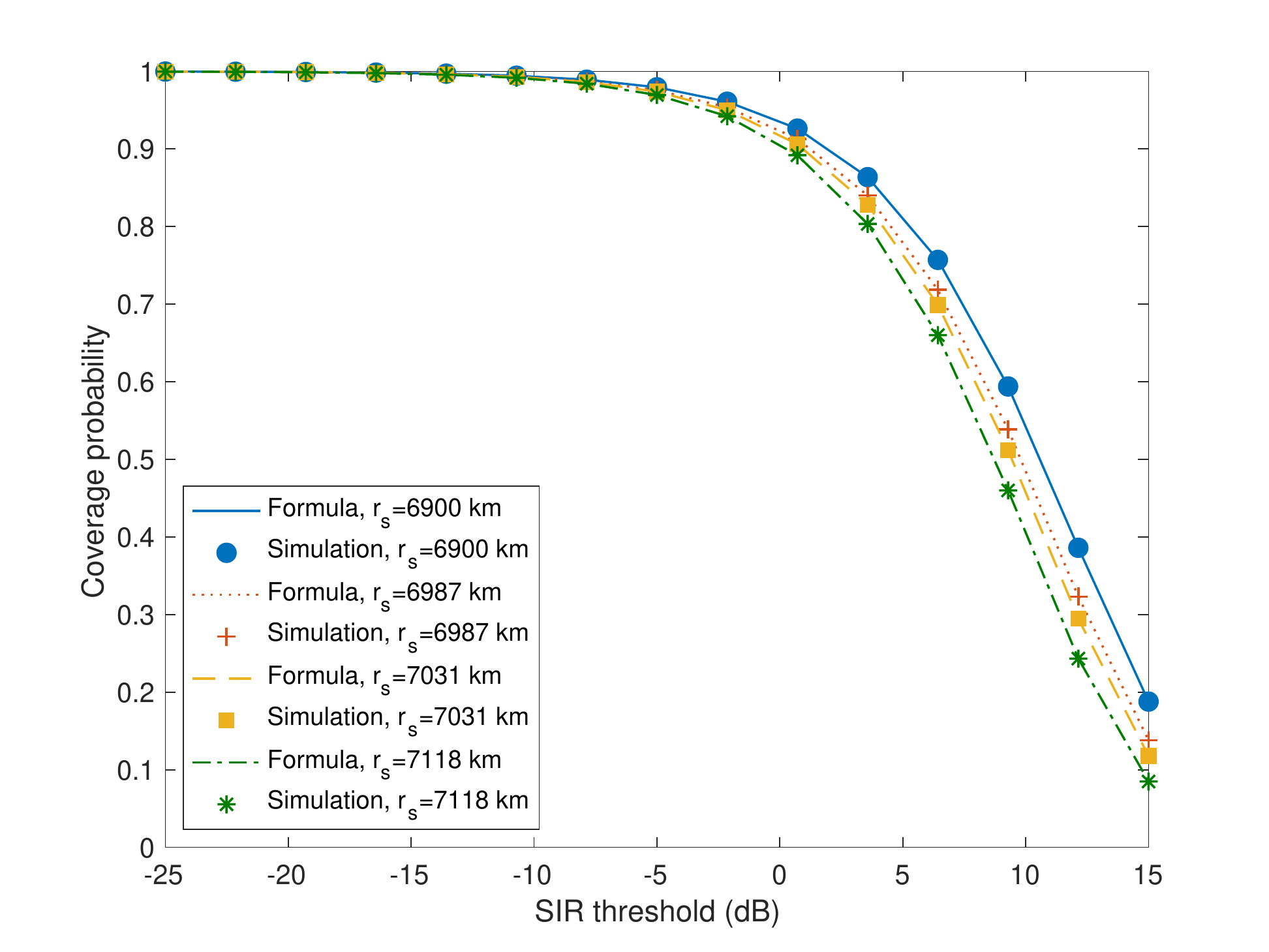}
	\caption{The coverage probability for various satellite altitudes. }
	\label{fig:rtausir}
\end{figure}

\begin{figure}
	\centering
	\includegraphics[width=.9\linewidth]{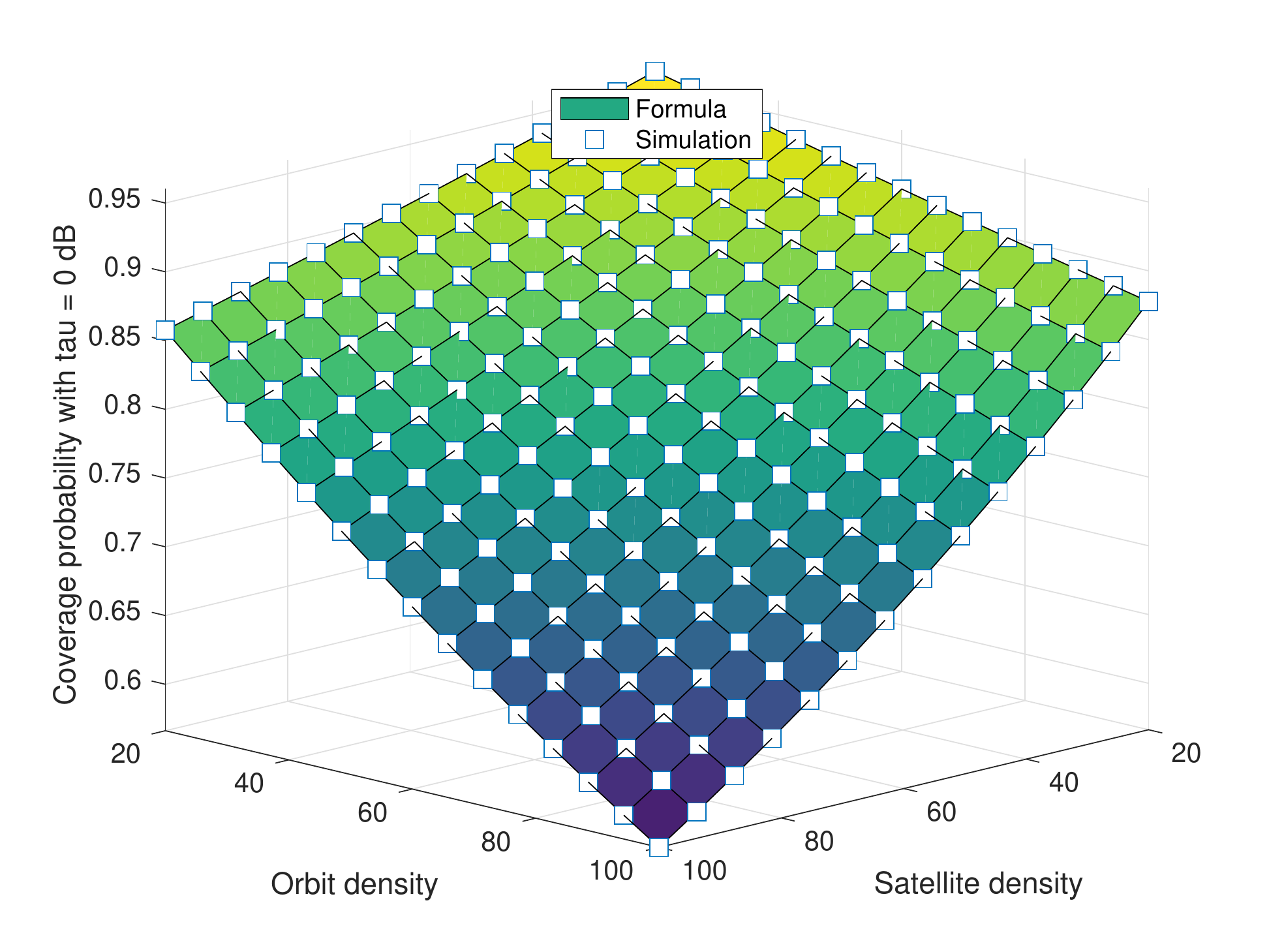}
	\caption{The $0$-dB coverage probability for various $\lambda$ and $\mu.$ }
	\label{fig:lambdamutau0}
\end{figure}

\begin{figure}
	\centering
	\includegraphics[width=.9\linewidth]{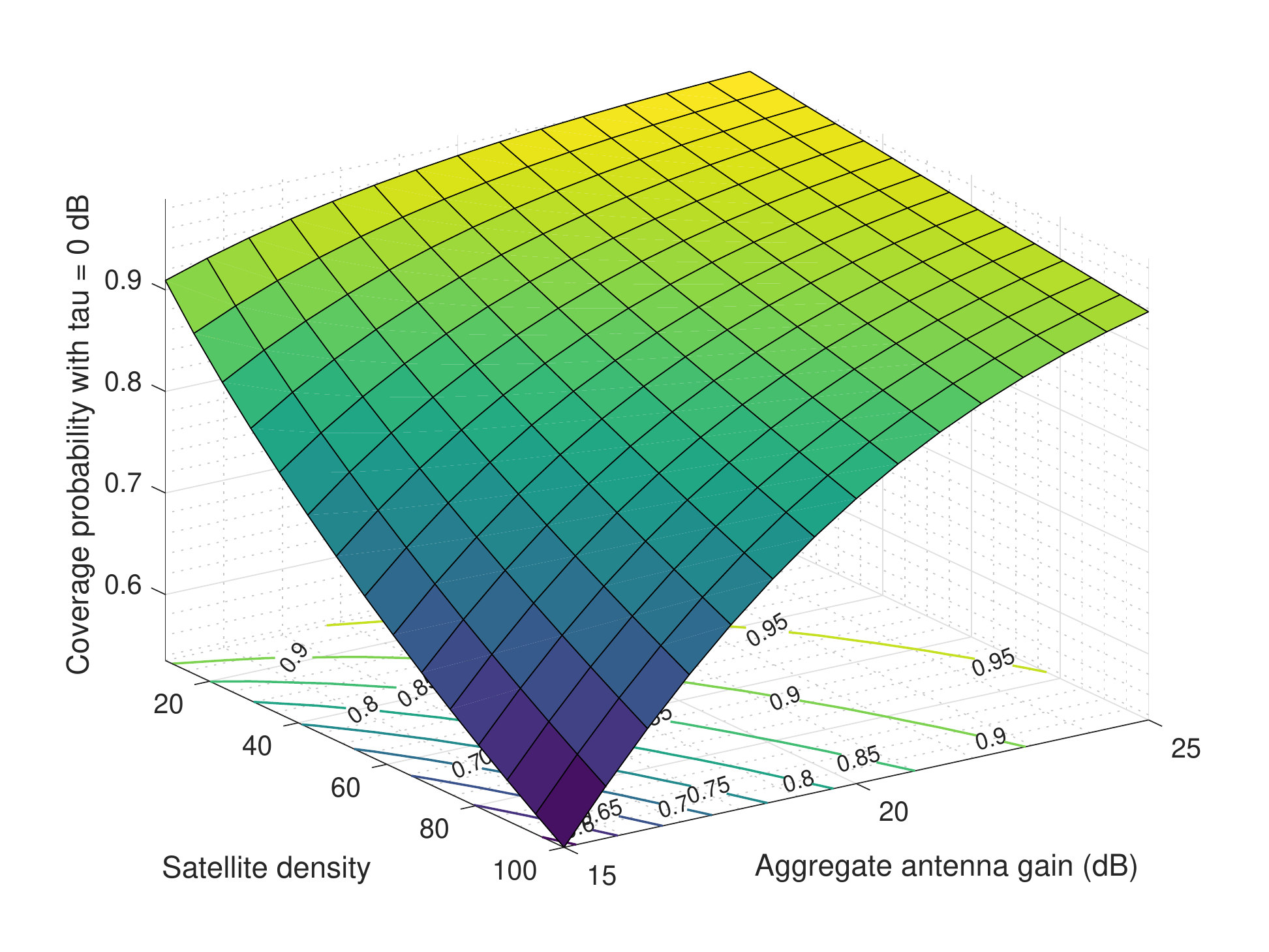}
	\caption{The $0$-dB coverage probability for various $\mu$ and $g$.}
	\label{fig:mugtau0}
\end{figure}

\begin{figure}
	\centering
	\includegraphics[width=1\linewidth]{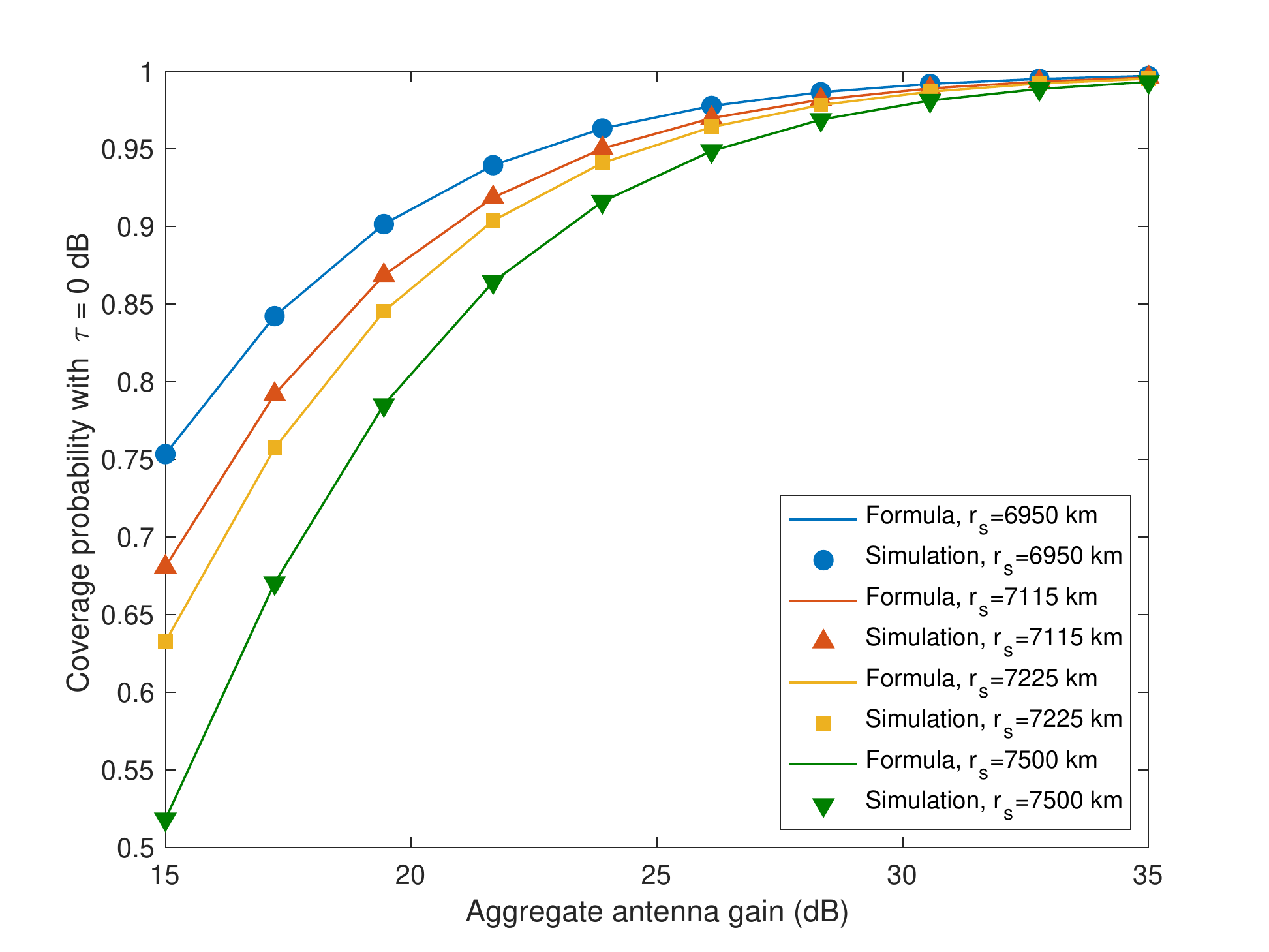}
	\caption{The $0$-dB coverage probability for various altitude. We use $\mu=40$.}
	\label{fig:rgsir0}
\end{figure}

Fig. \ref{fig:rtausir} displays the impact of satellite altitude on coverage probability. For the given $\lambda$ and $\mu$, it is observed that higher satellite altitudes typically lead to lower coverage probabilities. It is worth noting that the difference in coverage probability from various satellite altitudes is noticeable for higher thresholds.

Fig. \ref{fig:lambdamutau0} shows the $0$-dB coverage probability, namely the probability that the typical user has the SINR is greater than $0$ dB. The $0$-dB coverage probability of the typical user also indicates the total fraction of network users having their SINR greater than $0$ dB. Such an interpretation leads to important insights for satellite network design or its optimization. For instance, for $\lambda=50$ and $\mu=50$, about 85\% of users have SINR greater than $0$ dB. For $\lambda=100$ and $\mu=50$ or $\lambda=50$ and $\mu=100$, where the total average numbers of satellites are the same, their $0$-dB coverage probabilities are almost identical, at $0.74$. A deployment plan of forthcoming LEO or MEO satellite constellation---such as adding orbits or adding satellites on each orbit---can be effectively evaluated by adapting the proposed framework, offering a comprehensive tool to design LEO or MEO satellite networks for network operators. 

Fig. \ref{fig:mugtau0} gives the $0$-dB coverage probability for various $\mu$ and $g$. Here, we consider $g\in(15,25) $ dB. In practice, LEO or MEO satellite network operators deploy more satellites to cope with an increased number of ground users. Yet, this may lead to increased interference or decreased coverage due to the additional satellites. Occupying additional orbital planes may not be a feasible option, and in this case, a viable solution may be to deploy new LEO or MEO satellites with better antenna gain. Using the derived formula, Fig. \ref{fig:mugtau0} shows how much antenna gain from the LEO or MEO satellite is required to achieve the same coverage probability level for various densities of $\mu$. For instance, if the satellite density $\mu$ changes from $\mu_1$ to $\mu_2$, the network operator may also need additional antenna gain from $g$ to $g'$ to ensure the same quality of service for network users. In the same vein, we conduct a similar experiment in Fig. \ref{fig:rgsir0} where we examine the coverage benefit of having additional antenna gain for satellite altitudes.

	\begin{remark}\label{Remark2}
	\cc{In the evaluation of the coverage probability in Theorem \ref{Theorem:3}, we assume that the interference power dominates the noise power at the typical user.  For the values in Table \ref{Table:1}, we have
$\bE[I]\gg kT B_w $ namely $\sum_{X_i \in \Psi \setminus X_\star} pg_r\|X_i\|^{-\alpha} \gg kT B_w.$
		For the values of $\lambda$ and $\mu$ in Table \ref{Table:1}, the average number of visible satellites is 35. The typical user is associated with the nearest satellite whereas the remaining 34 satellites are creating interference to the typical user. }
\end{remark}

\begin{corollary}
	\cc{Accounting for the noise at the user, the coverage probability is given by 
	\begin{equation}
		\bP(\SINR>\tau) = e^{-\frac{kTB_w\tau}{pg}}\bE\left[\cL_{I}\left(\frac{\tau \|X_\star-n\|^{\alpha}} {pg}\right)\right]\label{36}
	\end{equation}
	where the second expectation of Eq. \eqref{36} is given by  Eq. \eqref{eq:theorem:4a}. In other words, the coverage probability with thermal noise is equal to the expression of the SIR coverage probability of Eq. \eqref{36} multiplied by the following constant:  $e^{-kTB_w\tau/(pg)}.$ }
\par 	\cc{The achievable rate or the ergodic capacity of downlink LEO or MEO satellite communications is given by  Eq. \eqref{eq:theorem:4.1}.}
\end{corollary}

\begin{IEEEproof}
	Since the noise is a constant, we obtain the result by by using Eq. \eqref{eq:166}. The achievable rate is  
		\begin{align}
				\text{Throughput} &= \int_0^\infty \bP(\log_2(1+\SINR)>r)\diff r \nnb\\
				&= \int_0^\infty \bP(\SINR>2^r-1)\diff r, 
			\end{align} 
		where we use the fact that $\log_2(1+\SINR)$ is a positive random variable. 
\end{IEEEproof}

\begin{figure*}
\vspace{-4mm}
			\begin{align}
		&\sum_{k=0}^{m-1}\frac{\lambda}{2 k!}\int\limits_{r_s-r_e}^{\sqrt{r_s^2-r_e^2}}\int\limits_{\pi/2-\xi}^{\pi/2+\xi}\!\bE_{{\Xi}}\!\!\left[ \left\{(-s)^k\frac{\partial^k}{\partial s^k}\left(h_1(z,v,s) \hspace{-5mm}\prod^{0<|\phi_i-\pi/2|<\xi}_{(\theta_{i},\phi_i)\in\Xi}\hspace{-5mm}h_{1}(z,\phi_i,s)\hspace{-5mm}\prod^{\xi<|\phi_i-\pi/2|<\overline{\varphi}}_{(\theta_{i},\phi_i)\in\Xi}\hspace{-5mm}h_{2}(z,\phi_i,s) \right)\right\}_{s=\frac{m\tau z^{\alpha}}{pg}}\right.\nnb\\
		&\hspace{48mm}\left.\frac{\mu u g(z,v)  \prod\limits_{(\theta_{i},\phi_i)\in\Xi}^{|\phi_i-\pi/2|<\xi}g(z,\phi_i)}{\pi r_e r_s\sqrt{1-{\cos^2(\xi)}{\csc^2(v)}}}  \right]\diff v\diff z \label{eq:theorem5}.
	\end{align}
	\rule{1\linewidth}{0.2mm}
\end{figure*}
	In below, we derive the coverage probability of the typical user for a general fading with $ m=1,2,...  $
	\begin{theorem}\label{Theorem:3}
		In the interference-limited regime with general $ m,$ the coverage probability of the typical user is given by Eq. \eqref{eq:theorem5} where $\xi,$  $\omega_{i,1}(z)$, $ \omega_{i,2}$, $\widetilde{K_{i,\omega}}$ are given by Eqs. \eqref{25},  \eqref{26}, \eqref{26.1}, and \eqref{26.2}, respectively. We also have 
				\begin{align}
			h_{1}(z,\phi_i,s)&{=}\exp\left({-\frac{\mu}{\pi}\int_{\omega_{i,1}(z)}^{\omega_{i,2}} 1 - \cL_H(sp/\widetilde{K_{i,\omega}}^\alpha)} \diff \omega\right),\nnb\\
			h_{2}(z,\phi_i,s)&{=}\exp\left({-\frac{\mu}{\pi}\int_{0}^{\omega_{i,2}} 1 - \cL_H(sp/\widetilde{K_{i,\omega}}^\alpha)} \diff \omega\right),\nnb\\
			g(z,\phi) &{=}\exp\left({-\frac{\mu}{\pi}\arcsin\left(\sqrt{1-\cos^2(\xi(z))\csc^2(\phi)}\right)} \right).\nnb
		\end{align}
	\end{theorem}
	\begin{IEEEproof} As in the proof of Theorem \ref{mainT}, in the interference-limited regime, the coverage probability of the network is 
		\begin{align}
			&\bP(\SINR>\tau)\nnb\\
			&= \bE_{\Xi}\left[\bP\left(\left.H>\frac{\tau I\|X_\star-n\|^{\alpha}}{ p g }\right|\Xi\right)\right]\nnb\\
			&=\bE_{\Xi,l_\star,D,I}\left.\left.\left.\left[\left.e^{-\frac{m\tau I z^\alpha}{p g}}\sum_{k=0}^{m-1}\frac{\frac{m^k\tau^k I^k z^{k\alpha}}{p^kg^k}}{k!}\right| D, l_\star \Xi \right.\right]\right]\right]\nnb\\
			&=\sum_{k=0}^{m-1}\frac{1}{k!}\bE_{\Xi,l_\star,D}\left.\!\left.\left[\left.s^k(-1)^k\frac{\partial^k}{\partial s^k}\cL_{I|D,l_\star,\Xi }(s)\right|_{s=\frac{m\tau z^\alpha}{pg}}\right.\right.\right]\nnb,
		\end{align}
		where  we used the probability distribution function of the Nakagami-$ m $ random variable. We also used the fact that the $ k $-th moment of the interference can be obtained by taking the $ k $-order derivative of the Laplace transform of the interference. Hence, we have 
		\begin{align}
			\bE_I\!\!\left[\left.e^{-\frac{m\tau I z^\alpha}{p g}}\sum_{k=0}^{m-1}\frac{\frac{m^k\tau^k I^k z^{k\alpha}}{p^kg^k}}{k!}\right| \cE \right]\!
\!			&=\!\!\left.\sum_{k=0}^{m-1}\frac{(-s)^k}{k!}\frac{\partial^k}{\partial s^k}\cL_{I|\cE}(s)\right.\nnb,
		\end{align}
	where ${s=\frac{m\tau z^\alpha}{pg}}$. As in the proof of Theorem \ref{mainT}, we have 
					\begin{align}
	\cL_{I|\cE}(s)&=\exp\left({-\frac{\mu}{\pi}
		\int_{\omega_{1}(z)}^{\omega_{2}} 1 - \cL_H(sp\widetilde{K_{\omega}}^{-\alpha})} \diff\omega\right)\nnb\\
		&\hspace{-5mm}\prod_{i=1,N}^{|\phi_{i}-\pi/2|<{\xi(z)}}\!\exp\left(\!{-\frac{\mu}{\pi}\int_{\omega_{i,1}(z)}^{\omega_{i,2}} 1 - \cL_H(sp/\widetilde{K_{i,\omega}}^\alpha)} \diff \omega\!\right)\nnb\\
	&\hspace{-5mm}\prod_{i=1,N}^{\xi(z)<|\phi_{i}-\pi/2|<\overline{\varphi}}\!\!\exp\left(\!{-\frac{\mu}{\pi}\int_{0}^{\omega_{i,2}} 1 - \cL_H(sp/\widetilde{K_{i,\omega}}^\alpha)} \diff \omega \!\right),
\end{align}
	where $\cL_H(s)$ is the Laplace transform of the random variable $H.$ The variables $\xi$,  $\omega_{i,1}(z)$, $\omega_{i,2}$, and $\widetilde{K_{i,\omega}}$ are given by Eqs. \eqref{25},  \eqref{26}, \eqref{26.1}, and \eqref{26.2}.
		\par Conditionally on $ \Xi $ and $l_\star, $ the PDF of $ D $ is given by taking the derivative of the CDF of $D$ as follows: 
		\begin{align}
			f_D(z)=&\frac{\mu z |\csc(\phi)|e^{-\frac{\mu}{\pi}\arcsin\left(\sqrt{1-\cos^2(\xi(z))\csc^2(\phi)}\right)} }{\pi r_er_s\sqrt{1-{\cos^2(\xi(z))}{\csc^2(\phi)}}}\nnb\\
			& \prod\limits^{|\phi_i-\pi/2|<\xi(z)}_{i=1,N}\!\!\!\!e^{-\frac{\mu}{\pi}\arcsin\left(\sqrt{1-\cos^2(\xi(z))\csc^2(\phi_i)}\right)}.
		\end{align}
		\par Finally, we get the result by combining all expressions and following the same steps as in the proof of Theorem \ref{mainT}.  
	\end{IEEEproof}

\section{Discussion}\label{S:6}

\subsection{Cox Model to a Forthcoming Constellation}
\cc{In this section, we focus on describing the Starlink constellation second-generation (2A) leveraging the proposed Cox point process by comparing the SINR coverage probability of the Starlink and the Cox model. Note that the Starlink constellation 2A is a forthcoming plan and it is an example among many future satellite constellations. Therefore, the aim here is to demonstrate that the proposed model can be used to model a planned constellation. Since the local distribution of the Starlink constellation varies with user latitudes, we develop a local approximation technique based on a moment matching. Specifically, we adjust the orbital and satellite parameters, denoted by $\lambda$ and $\mu$, to ensure that the Cox constellation and the Starlink constellation have, on average, the same number of satellites.}

\cc{For the upcoming Starlink constellation, we refer to the Starlink deployment plan available at the FCC \cite{FCCKuiper}, which involves $28$ orbital planes at an altitude of $525$ km and an inclination of $43$ degrees, $28 $ orbital planes at $530 $ km with an inclination of $53$ degrees, and $28$ planes at an altitude of $535$ km with an inclination of $33$ degrees. Each plane will accommodate $120$ satellites. To numerically derive the network performance of this constellation, we consider a frequency reuse factor of $4$ where $30$ satellites on each orbit use the same spectrum resource. }

\begin{figure}
 	\centering
 	\includegraphics[width=.9\linewidth]{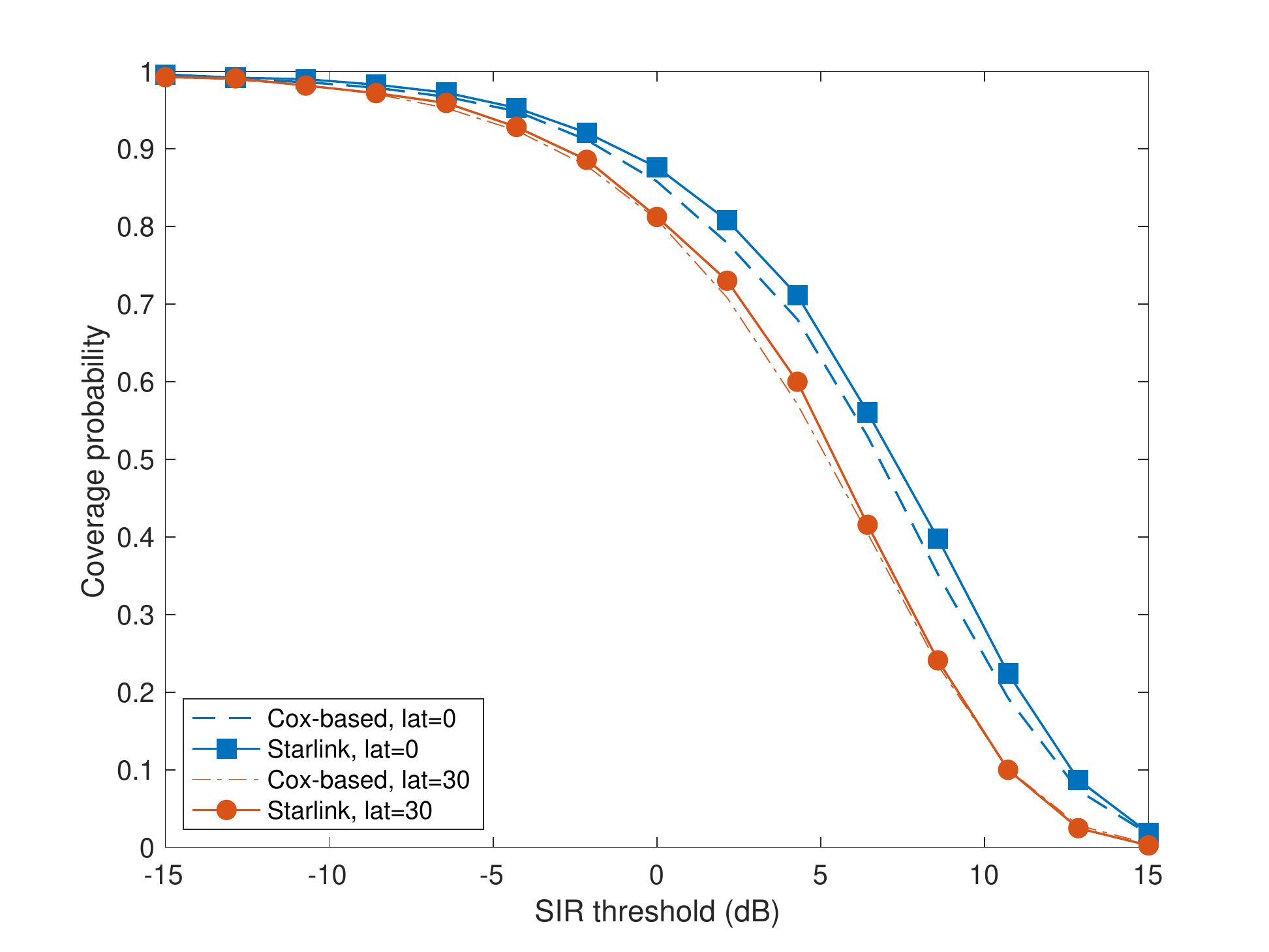}
 	\caption{The coverage probability of Starlink and Cox constellations for users at latitudes of $0$ and $30$ degrees. $r_s=7000$. }
 	\label{fig:slvscoxtotal}
 \end{figure}

\cc{Fig. \ref{fig:slvscoxtotal} illustrates the coverage probability based on the upcoming Starlink constellation and the proposed Cox point process. Note that in order to numerically evaluate the SIR coverage probability of the typical user in the Starlink constellation, we simulate $10^6$ satellite layouts according to the available information above. Then, the SIR of the typical user is obtained for each layout and the collected SIR values are finally combined together as the CCDF of the SIR random variable. Based on the moment matching method specified above, we find the parameter settings $(\lambda,\mu) = (38,80)$ for the latitude of $30 $ degrees and $(\lambda,\mu)=(100,21)  $  for the latitude of $0$ degree. From Fig. \ref{fig:slvscoxtotal}, we observe that at the latitude of $30$ degrees, the proposed Cox point process accurately emulates the constellation. Similarly, we observe that at a latitude of $0$ degree, the proposed Cox point process replicates a coverage probability comparable to that of the Starlink constellation, with differences of less than $0.5$ dB. The marginal difference arises from the geometric distinction that while the developed Cox point process is isotropic, the satellites of the forthcoming Starlink 2A constellation are regularly separated on each orbit and the orbits' inclinations are having only three values at $43,53,33$ degrees. }

\subsection{Cox model and Binomial Model}

\cc{Here, we highlight the difference between our Cox point process and the binomial point process. Analytical stochastic geometry models, such as the binomial model \cite{9079921,9177073,9218989,9497773,9678973,9861782} or our Cox model, approximate existing or forthcoming target constellations. For example, the binomial point process features a geometric parameter, denoted as $N$, representing the number of satellites. This parameter $N$ is then fine-tuned to match the mean number of visible satellites of the target constellation. In contrast, the proposed Cox point process features two parameters, $\lambda$ and $\mu$, and these two parameters are adjusted jointly to match the mean number of orbits and the mean number of satellites at the same time. Consequently, the presence of two parameters allows one to approximate the higher-order geometric characteristics of the target constellation, offering a better representation of orbital planes that LEO or MEO satellite constellations feature in space}. 

\cc{It is important to note that a proper comparison must be based on various mathematical properties of those point processes such as the repulsion between points often examined by the Ripley-K function of the point process. The detailed comparison between binomial and Cox is left for future work. Here, we simply study them by underlining their differences in performance metrics such as the no-satellite probability or the coverage probability.}  

\cc{First, we compare the no-satellite probability of the proposed Cox point process to the no-satellite probability of the binomial point process. Consider the total numbers of satellites to be $N=100,200,400$, and $ 1000$, all at the altitudes of $550$ km. For the binomial model, we empirically compute the no-satellite probability by first creating the binomial satellite point process and then counting each event that there is no satellite point visible from the typical observer $(0,0,r_e)$. We obtain the no-satellite probabilities for $N=100,200,400$, and $ 1000$ as follows: $0.0176$ for $N=100$ and $0.0003$ for $N=200$. For $N=400$ and $N=1000$, the no-satellite probabilities are negligible.	}

\cc{On the other hand, the no-satellite probability of the proposed Cox point process is given by Eq. \eqref{eq:Theorem2}. For a mean of $100$ satellites, we basically have numerous ways to distribute the mean of $100$ satellites by varying the parameters $\lambda$ and $\mu$ together, namely $100=\lambda\mu$. For instance, we can use the following 10 cases: $(\lambda,\mu)=(10,100/10),(20,100/20), \ldots,$ and $(100,100/100)$. The first case indicates that there are $100$ satellites distributed over $10$ distinct orbits whereas the last case indicates the $100$ satellites are distributed over $100$ distinct orbits. For the above $10 $ cases, we numerically obtain the following no-satellite probabilities: $0.0886, 0.0471, \ldots$, and  $0.0235$. It is worth noting that the first and last cases created by the Cox model are topologically very different because of the existence of the orbit parameter $\lambda$. We observe that the proposed Cox model can reproduce an accurate no-satellite probability relevant to each deployment scenario in practice, by delineating various deployment scenarios such as a dense orbit scenario with few orbits and many satellites per orbit or a sparse orbit scenario with many orbits and few satellites per orbit. Note, on the other hand, that the binomial model produces a single value of $0.0176.$ } 

\cc{Similarly, for a mean of $200$ satellites, the no-satellite probability of the binomial model is very low at $0.0003$. On the other hand, for the proposed Cox model, we can vary $(\lambda,\mu)=(10,200/10),(20,200/20), \ldots,$ and $(200,200/200)$. The no-satellite probabilities are given by $0.033, 0.0079,\ldots,$ and $0.0006$. Consider a deployment scenario where $200$ satellites are uniformly distributed over $10$ isotropic orbits. The proposed Cox model produces an accurate estimate of the no-satellite probability as $0.033$, whereas the binomial model produces an estimate of the no-satellite probability value as $0.0003$, which is twice larger than the extreme case that the Cox model produces with even $200 $ orbits.  This low no-satellite probability of the binomial model stems from the fact that the points of the binomial point process are independent and uniformly distributed over the sphere. Therefore, it is statistically enforced to produce an extremely low no-satellite probability. In conclusion, unlike the existing binomial point process where only a single layout is created for a given number of satellites, the developed Cox model can reproduce various deployment scenarios. }

\begin{figure}
	\centering
	\includegraphics[width=.9\linewidth]{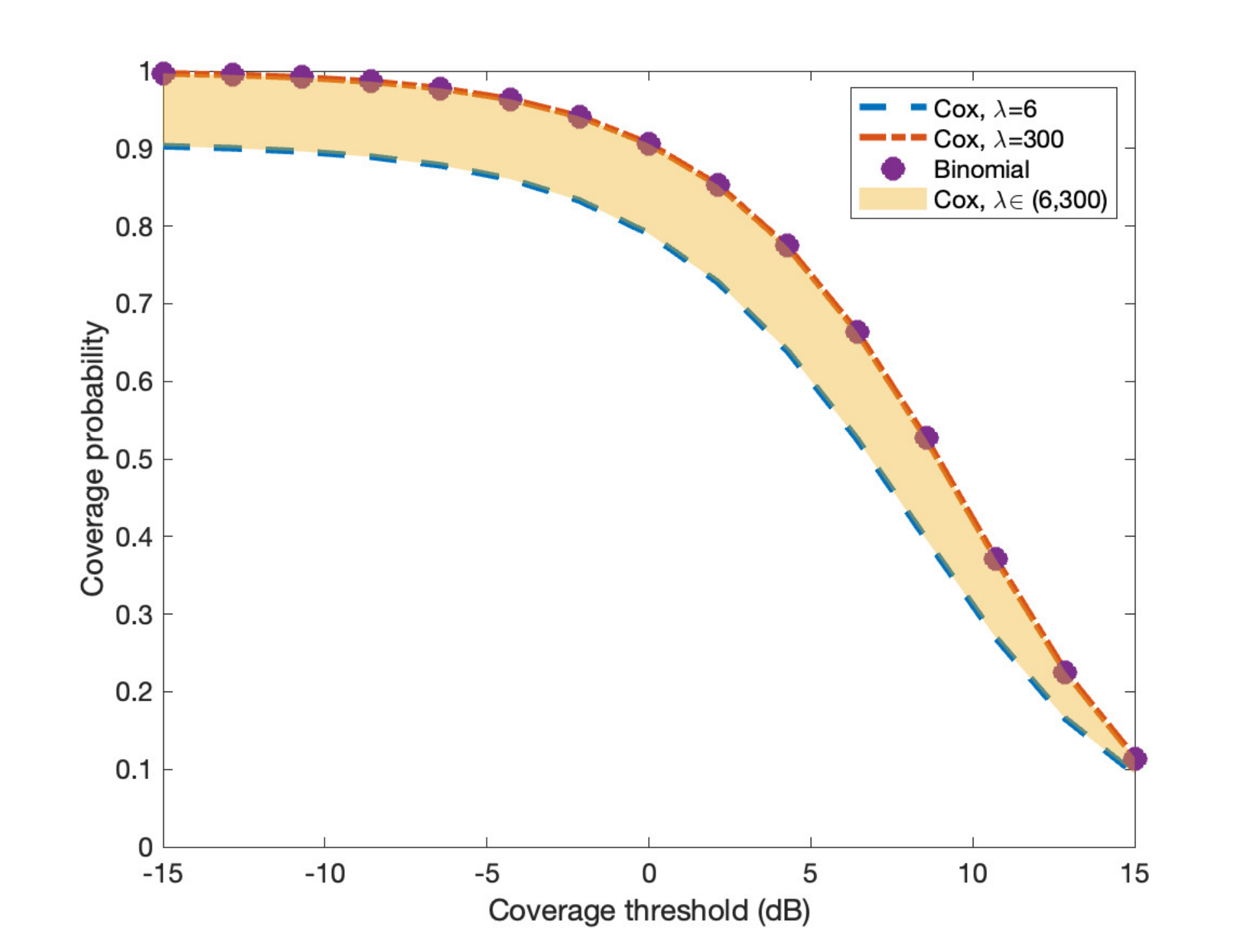}
	\caption{Our Cox model features a wide range of coverage probabilities whereas a single coverage probability curve appears for the binomial model. }
	\label{fig:section62}
\end{figure}

\cc{Secondly, in Fig. \ref{fig:section62} we numerically compare the SIR coverage probabilities of both the binomial and Cox satellite point processes. We consider a single target constellation having $300$ satellites at an altitude of $550$ km and the rest of simulation parameters are given in Table \ref{Table:1}. The modeling technique based on the binomial point process results in a single curve for the coverage probability of the typical user. On the other hand, the modeling based on Cox point process has additional flexibility through the number of orbits. This allows for a wide range of achievable coverage probabilities by varying $\lambda$ and $\mu$. Thanks to the flexibility, we observe that even when the two point processes have the same mean number of satellites, the achievable region of the coverage probabilities derived under the Cox point process may include the coverage probability derived under the binomial point process. In Fig. \ref{fig:section62}, the shaded area is the coverage probabilities produced by the Cox model for a single mean number of satellites. For this mean number, the Cox point process produces a continuum of orbital planes by varying the orbit parameter $\lambda$ from a very small number to a very large number. From the wide range of coverage probabilities that the Cox model produces, Fig. \ref{fig:section62} suggests that the Cox model may include the binomial model in terms of the coverage probability.} 


\subsection{Direct Orbit Representation}
In this paper, we consider a non-homogeneous density function ${\lambda\sin(\phi)}/{(2\pi)}$ to create the orbits that are isotropic and invariant by rotations. Some forthcoming constellations may exhibit regularly spaced orbits where orbits' longitudes are regularly spaced while their inclinations are fixed to some values. For such cases, we modify the density function to directly represent the constellation. For instance, the satellite constellation with regularly spaced $N$ orbits with an inclination angle of 43 degrees (0.75 rad) are represented as follows: $\bar{\Xi} = \sum_{i=1}^N \delta_{\frac{\pi}{N}(i-1)+\bar{U}, 0.75}$ where $\bar{U}$ is a uniform random variable in the interval $(0,\pi/N)$. Each point of $\bar{\Xi}$ represents an orbit of inclination $0.75$ rad and of longitude $\{(i-1)\pi/N\}_{i=1,\ldots,N}$. The uniform random shift is required in the above expression  since it ensures that the longitudinally rotated version of $\bar{\Xi}$ and its original ${\Xi}$ have the same distribution. In other words, $\bar{\Xi}$ is longitudinally rotation invariant. The above representation exactly captures the regularly spaced single constellation, and therefore, leveraging the machinery that we explored in this paper, one can also derive the no-satellite probability or the SIR coverage probability of the regularly spaced satellite network.

Another example of regularly spaced LEO or MEO satellite constellation is a Walker constellation.  A simple Walker constellation exhibits $N$ orbits each with the inclination of $\pi/2$  degrees; $\tilde{\Xi} = \sum_{i=1}^N \delta_{\frac{\pi}{N}(i-1)+\tilde{U}, \pi/2}$
where $\tilde{U}$ is a uniform random variable in the interval $(0,\pi/N)$. 

Nevertheless, it is essential to note that since the above orbit processes are longitudinally rotation invariant, the probability law of the orbit process will alter as we rotate the orbit process along any axes other than the $z$-axis. In other words, observers at different latitudes will see different satellite distributions, and therefore, the no-satellite probability, the interference, and the SIR coverage probability must depend on the latitude of the typical observer. A more detailed analysis based on the modified orbit process is left for future work.

	\section{Conclusion}

We have constructed an analytical framework for modeling and analyzing downlink LEO or MEO satellite communications. By developing an isotropic Cox point process for LEO or MEO satellites, we introduced a novel method for incorporating the essential geometry of LEO or MEO satellite constellations, where satellites are always located on orbits. We calculate the no-satellite probability for the typical user. Assuming that network users receive their downlink signals from their closest satellites, we evaluate the Laplace transform of the interference and assess the coverage probability for the typical user. The derived metrics represent network performance as functions of key network parameters. Thanks to isotropy, our analysis of the typical user's network performance can be interpreted as the network performance spatially averaged over all users across the network. Therefore, our analysis can be applied to the systematic design of large-scale downlink satellite communication networks. We demonstrate that our Cox point process effectively replicates a forthcoming LEO or MEO satellite constellation through moment matching approximation. Additionally, our Cox model provides numerous ways to model real or forthcoming constellations, in contrast to models based on binomial or Poisson point processes.

Future work will concentrate on addressing the limitations of the proposed model. For example, our framework currently assumes isotropic orbits, and a similar approach can be used to explore and investigate non-isotropic orbits. Similarly, the developed isotropic orbit may feature regularly-spaced satellites exclusively located on each orbit. Moreover, the proposed model assumes a fixed altitude for satellites. A similar approach can also be used to analyze satellite networks at various altitudes.


 \appendices
	\section{}\label{length of an arc}
	\begin{figure}
		\centering
		\includegraphics[width=.9\linewidth]{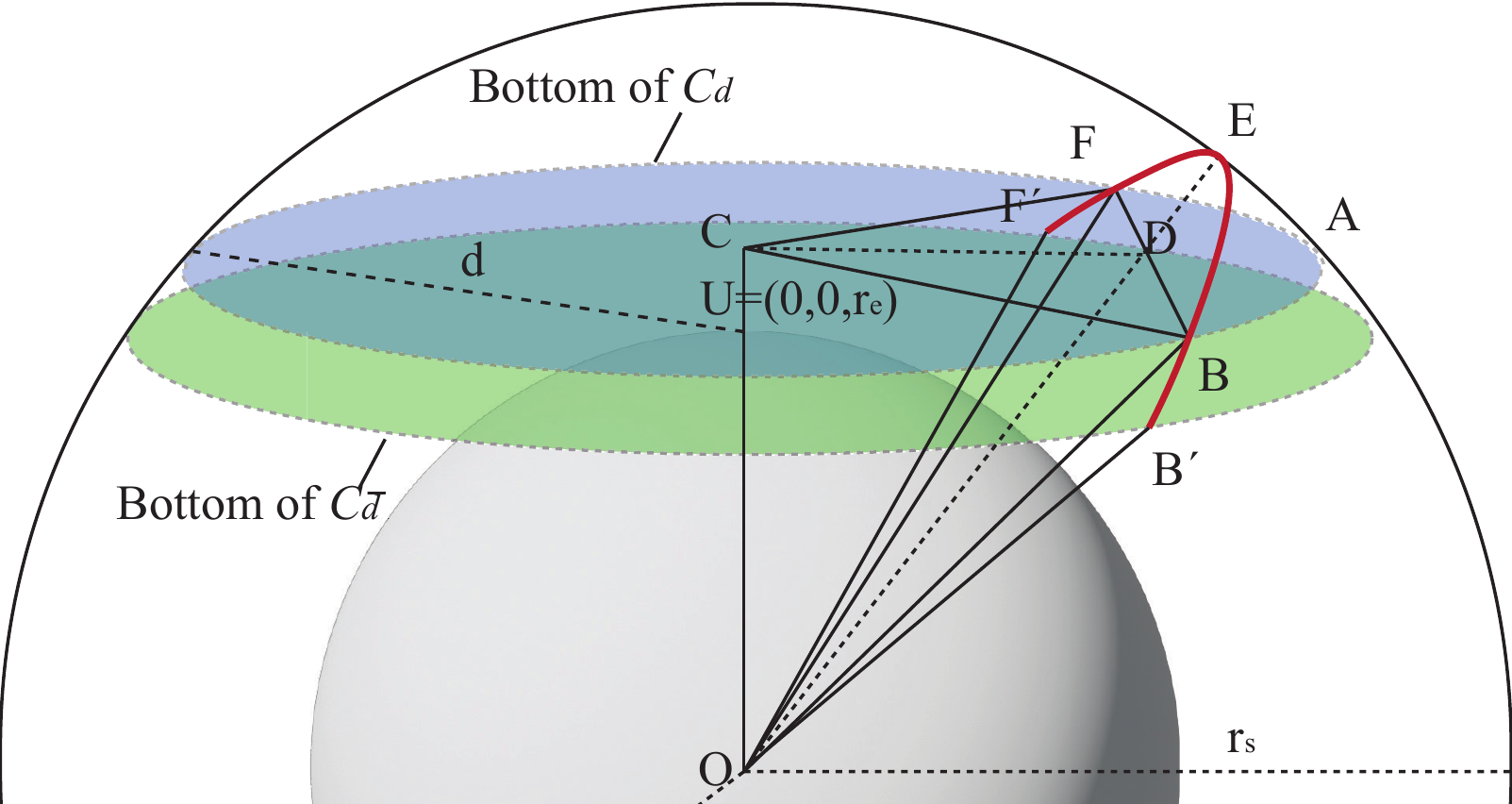} 
		\caption{Illustration of spherical caps $ C_d $ and $ C_{\overline{d}} $ where $  r_s-r_e<d<\overline{d} $ where $\overline{d}$.}
		\label{fig:figure1}
	\end{figure}
	
	Let us treat $ d $ as a constant between $ r_s-r_e $ and $ \sqrt{r_s^2-r_e^2} $ for the moment. Fig. \ref{fig:figure1} shows the bottom of spherical cap $ C_d $ where the arc $ \arc{BF} $  is defined as $ C_{d}\cap l(\theta,\varphi) $.
	\par For a triangle $ \triangle AOU,  $ in Fig. \ref{fig:figure1} $ \overline{AO}  = r_s$ and let $ \xi $ be the angle $ \angle AOU $. Then, we have  $\cos(\xi) = \frac{r_s^2+r_e^2-d^2}{2r_er_s}.$

	\par For $ \triangle BOC, $ $ \angle BOC = \xi$  and 
 we have $ \overline{OB}=r_s, \overline{OC} = r_s\cos(\xi), \overline{BC}=r_s\sin(\xi) $. Then, since the orbital plane of $ l(\theta,\phi) $ has the inclination of $ \phi $,  $ \angle DOC = \pi/2-\phi $, $ \overline{OC} = r_s\cos(\xi), $ and $ \overline{CD} = r_s\cos(\xi)\cot(\phi). $
 
 \par For $ \triangle BCD, $ we have $ \overline{BC} = r_s\sin(\xi)$ and $ \overline{CD}=r_s\cos(\xi)\cot(\phi) $. Since $ \angle BDC = \pi/2, $ we have $ \overline{BD} = \sqrt{r_s^2\sin^2(\xi) - r_s^2 \cos^2(\xi)\cot^2(\phi)}. $ 
\par Now consider a trangle $ \triangle OBD $ in the orbital plane. We have $ \overline{OB} = r_s $ and $ \overline{BD} = \sqrt{r_s^2\sin^2(\xi) - r_s^2\cos^2(\xi)\cot^2(\phi)}.  $ Let $ {\kappa}' = \angle BOD $. Since $ \angle BDO =\pi/2$, we get
\begin{equation}
	\sin(\kappa')=\frac{\overline{BD}}{r_s}=  \sqrt{\sin^2(\xi) - \cos^2(\xi)\cot^2(\phi)}.
\end{equation}
\par As a result, for $\pi/2-\xi<\phi<\pi/2+\xi$  the length of the arc $ \arc{BF} $ is  $2r_s\arcsin(\sqrt{1-\cos^2(\xi)\csc^2(\phi)}).$


\section{}\label{A:2}
Based on the simple geometry, the $(x,y,z)$ coordinates of the satellite with orbital angle $\omega_j$ on the orbit $ l(\theta_{i},\phi_{i}) $ are 
\begin{align}
	x &= \sqrt{r_s^2\cos^2(\omega_j)+ r_s^2\sin^2(\omega_j)\cos^2(\phi_{i})}\cos\left(\tilde{\theta_{i}}+\theta\right),\nnb\\
	y &= \sqrt{r_s^2\cos^2(\omega_j)+ r_s^2\sin^2(\omega_j)\cos^2(\phi_{i})}\sin\left(\tilde{\theta_{i}}+\theta\right),\nnb\\
	z &= r_s\sin(\omega_j)\sin(\phi_{i}),\nnb\\
	\tilde{\theta} &= \arctan\left({\tan(\omega_j)\cos(\phi_{i})}\right).\nnb
\end{align}
Moreover,  the distance from $X_{j,i} $ to $n$ is given by 
\begin{align}
	\|X_{j,i}-n\| \! =\sqrt{r_s^2-2r_s r_e\sin(\omega_j)\sin({\phi}_i)+r_e^2}=K_{i,j}.\label{kfunction}
\end{align} 

\section*{Acknowledgment}
This paper is a joint study with the support of Korea-France Cooperative Development Program (STAR) from the NRF, funded by the Korean government in the year 2024 (RS-2024-00247682). The work of Chang-Sik Choi was supported by the NRF-RS-2024-00334240. The work of Francois Baccelli was supported by ERC NEMO grant 788851 to INRIA and by the French National Agency for Research (ANR) via the project n°ANR-22-PEFT-0010 of the France 2030 program PEPR.
{\em Réseaux du Futur}.

\bibliographystyle{IEEEtran}
\bibliography{ref}

\end{document}